\documentclass[useAMS,usenatbib]{mnras}
\usepackage{epsfig}
\usepackage{graphicx}
\usepackage{caption}
\usepackage{float}
\usepackage{amsmath}
\usepackage{threeparttable}
\usepackage{gensymb}
\usepackage{subfig}
\graphicspath{ {images/} }
\usepackage{kantlipsum}
\usepackage{etoolbox}
\usepackage{aas_macros}
\usepackage{color}
\usepackage{tikz}
\usepackage{hyperref}
\def\checkmark{\tikz\fill[scale=0.4](0,.35) -- (.25,0) -- (1,.7) -- (.25,.15) -- cycle;} 

\title[KiDS CNN lens finder]{Finding Strong Gravitational Lenses in the Kilo Degree Survey with Convolutional Neural Networks}

\def\Fig{\mbox{Fig.~}}

\def\Tab{\mbox{Table~}}
\def\Sec{\mbox{Sect.~}}
\def\Secs{\mbox{Sections~}}

\def\Eq{\mbox{Eq.~}}

\def\RE{\mbox{$R_{\rm E}$}}
\def\Mauto{\mbox{{\tt MAG\_AUTO}}}

\def\sigs{\mbox{$\sigma_{\rm \star}$}}
\def\sigSIS{\mbox{$\sigma_{\rm SIS}$}}
\def\Msun{\mbox{$M_\odot$}}
\def\mst{\mbox{$M_{\star}$}}
\def\Zsun{\mbox{$Z_{\odot}$}}

\def\lsim{\mathrel{\rlap{\lower3.5pt\hbox{\hskip0.5pt$\sim$}}
    \raise0.5pt\hbox{$<$}}}
\def\gsim{~\rlap{$>$}{\lower 1.0ex\hbox{$\sim$}}}

\author[C.~E.~Petrillo et al.]{C.~E.~Petrillo$^{1}$\thanks{E-mail: petrillo@astro.rug.nl},  C.~Tortora$^{1}$, S.~Chatterjee$^{1}$, G.~Vernardos$^{1}$, L.~V.~E.~Koopmans$^{1}$, \and G.~Verdoes~Kleijn$^{1}$, N.~R.~Napolitano$^{2}$, G.~Covone$^{3}$, P.~Schneider$^{4}$, A.~Grado$^{2}$, \and J.~McFarland$^{1}$\vspace{0.2cm} \\ 
$^{1}$Kapteyn Astronomical Institute, University of Groningen, Postbus 800, 9700 AV, Groningen, The Netherlands\\
$^{2}$INAF - Osservatorio Astronomico di Capodimonte, Salita Moiariello, 16, 80131 Napoli, Italy\\
$^{3}$Dipartimento di Scienze Fisiche, Universit\`a di Napoli Federico II, Compl. Univ. Monte S. Angelo, 80126 Napoli, Italy\\
$^{4}$Argelander-Institut f\"ur Astronomie, Auf dem H\"ugel 71, D-53121 Bonn, Germany}

\hypersetup{draft}
\begin{document}

\pagerange{\pageref{firstpage}--\pageref{lastpage}} \pubyear{2016}

\maketitle

\label{firstpage}

\begin{abstract}
The volume of data that will be produced by new-generation surveys requires automatic classification methods to select and analyze sources. Indeed, this is the case for the search for strong gravitational lenses, where the population of the detectable lensed sources is only a very small fraction of the full source population.
We apply for the first time a morphological classification method based on a Convolutional Neural Network (CNN) for recognizing strong gravitational lenses in $255$ square degrees of the Kilo Degree Survey (KiDS), one of the current-generation optical wide surveys. The CNN is currently optimized to recognize lenses with Einstein radii $\gsim 1.4$ arcsec, about twice the $r$-band seeing in KiDS.
In a sample of $21789$ colour-magnitude selected Luminous Red Galaxies (LRG), of which three are known lenses, the CNN retrieves 761 strong-lens candidates and correctly classifies two out of three of the known lenses.  The misclassified lens has an Einstein radius below the range on which the algorithm is trained. We down-select the most reliable 56 candidates by a joint visual inspection. This final sample is presented and discussed. A conservative estimate based on our results shows that with our proposed method it should be possible to find $\sim100$ massive LRG-galaxy lenses at $z\lsim 0.4$ in KiDS when completed. In the most optimistic scenario this number can grow considerably (to maximally $\sim$2400 lenses), when widening the colour-magnitude selection and training the CNN to recognize smaller image-separation lens systems.
\end{abstract}

\begin{keywords}
gravitational lensing: Strong -- methods: statistical -- galaxies: elliptical and lenticular, cD
\end{keywords}

\section{Introduction}
Strong gravitational lensing is a rare phenomenon which provides very tight constraints on the projected mass of the foreground lens galaxy. In fact, the total mass (dark plus baryonic) within the Einstein radius depends almost solely on the space-time geometry of the lensing system (the source and the lens redshift and the cosmological parameters). 
For this reason, strong lensing is a unique tool, if combined with central velocity dispersion measurements and stellar population analysis, to estimate the fraction of dark matter in the central regions of galaxy-scale halos (e.g., \citealt{gavazzi2007,jiang2007,grillo2010,cardone2009,Cardone2010,tortora2010central,more2011,ruff2011,sonnenfeld2015sl2s}), and to constrain the slope of the inner mass density profile (e.g., \citealt{Treu2002MNRAS,Treu2002,koopmans2006,koopmans2003,Moore2008,barnabe2009,koopmans2009,cao2016limits}). 

Gravitational lenses can be also used to constrain the stellar initial mass function (e.g., \citealt{treu2010, ferreras2010, Spiniello:2011p8239, brewer2012swells,sonnenfeld2015sl2s,posacki2015stellar,leier2016strong}) and to independently measure the Hubble constant through time delays (e.g., 
\citealt{suyu2010,bonvin2016}).
In addition, strong lensing gives magnified views of background objects otherwise inaccessible to observations (e.g., \citealt{impellizzeri2008,swinbank2009,richard2011,deane2013,treu2015grism,Mason2016}). 

A homogeneously selected large lens sample can improve dramatically the effectiveness of the methods and the reliability of the results from gravitational lensing studies. The largest homogeneous sample so far is provided by the Sloan Lens ACS Survey (SLACS; \citealt{bolton2008}) with almost 100 observed lenses. In the future, deep high resolution wide surveys have the potential to produce samples three orders of magnitude larger than the current known lenses. These large numbers will allow to, e.g., greatly improve the precision in the mass density slope measurements \citep{Barnab2011}, in better estimate the presence of substructure \citep{Vegetti2009} and to put constraints on the nature of dark matter \citep{Li2016}.

Upcoming telescopes, such as Euclid \citep{Laureijs:2011wi} and the Large Synoptic Survey Telescope (LSST; \citealt{abell2009}), will increase the rate of discovery of new lenses, reaching the number of $\sim 10^5$ new strong lensing systems \citep{oguri2010gravitationally,pawase2012,collet2015}. Also, the number of lenses that will be observed by the Square Kilometer Array is expected to be of the same of order of magnitude \citep{McKean2015}. The ongoing optical wide surveys, such as the Kilo Degree Survey (KiDS; see Sec. \ref{sec:KiDS}), the Dark Energy Survey (DES; \citealt{DES}) and the Subaru Hyper Suprime-Cam Survey (\citealt{HSCmiyazaki2012hyper}) are expected to find samples of lenses of the order of $\sim 10^3$ (see, e.g, \citealt{collet2015}). Sub-mm observations from Herschel \citep{negrello2010} and the South Pole Telescope (\citealt{carlstrom2011}), together with deeper, high resolution observations from the the Atacama Large Millimeter/sub-millimeter Array, are expected to provide several hundred new lenses as well.

Traditionally, the search of extended lens features (i.e., arcs and rings) relied heavily on the visual inspection of the targets. This is still the best approach for small samples of objects, but is impractical for the ongoing and new generation surveys given the large number of targets that need to be inspected. Accordingly, numerous automatic lens finders have been developed in recent years.  Most are based on the identification of arc-like shapes (e.g., \citealt{Lenzen2004,Horesh2005,Alard2006,Estrada2007,Seidel2007,Kubo2008,More2012}). The same approach, together with a colour selection, is employed by \cite{Maturi2014}. Another method consists of subtracting the light of the central galaxies using multiband images and then analyse the image residuals \citep{Gavazzi2014}. \cite{Joseph2014} follow a similar approach but employing machine-learning techniques to analyse single-band images. Instead \cite{Brault2015} model the probability that the targets are actual lenses. Very recently \cite{Bom2016} have developed an artificial neural network for recognizing strong lenses that uses as entries a set of morphological measurements of the targets. A completely different approach based on crouwdsourcing is employed in the Space Warps project \citep{Marshall2016,more2016}, with volunteers visually inspecting and classifying galaxy cutouts through a web applet\footnote{\href{https://spacewarps.org/}{\tt https://spacewarps.org/}}. All these automatic methods have their advantages and disadvantages and perform at their best with different typologies of lenses, quantity and kind of data available. A detailed comparison between these methods should be done on a common dataset, but is beyond the scope of this paper. 

Convolutional Neural Networks (CNNs; \citealt{fukushima1980neocognitron,lecun1998gradient}) are a state of the art class of machine learning algorithm particularly suitable for image recognition tasks. The ImageNet Large Scale Visual Recognition Competition (ILSVRC; \citealt{ILSVRC15}; the most important image classification competition) of the last four years has been won by groups utilizing CNNs. The advantage of CNNs with respect to other pattern recognition algorithms is that they automatically define and extract representative features from the images during the learning process.
Although the theoretical basis of CNNs was built in the the 1980s and the 1990s, only in the last years do CNNs generally outperform other algorithms due to to the advent of large labelled datasets, improved algorithms and faster training times on e.g. Graphics Processing Units (GPUs). We refer the interested reader to the reviews by \cite{schmidhuber2015deep}, \cite{lecun2015deep} and \cite{guo2016deep} for a detailed introduction to CNNs.

The first application of CNNs to astronomical data was made by \cite{Hala2014} for classifying spectra in the Sloan Digital Sky Survey (SDSS; \citealt{SDSS}). Then, \cite{dieleman2015rotation}
\footnote{The method won a challenge against other techniques \href{https://www.kaggle.com/c/galaxy-zoo-the-galaxy-challenge/}{\tt https://www.kaggle.com/c/galaxy-zoo-the-galaxy-challenge/}} 
used CNNs to morphological classify SDSS galaxies. Subsequently, \cite{Huertas2015} used the same set-up of \cite{dieleman2015rotation} for classifying the morphology of high-z galaxies from the Cosmic Assembly Near-IR Deep Extragalactic Legacy Survey \citep{CANDELS}. More recently, \cite{Hoyle2016} adopted CNNs for estimating photometric redshifts of SDSS galaxies. CNNs have been employed also by \cite{Kim2016} for star/galaxy classification.

In this paper we present our morphological lens-finder which is based on CNNs. We apply it to the third data release of KiDS \citep{deJong+15_KiDS_paperI, deJong+17_KiDS_DR3}, starting a systematic census of strong lenses. This project, which consists of both visual and automatic inspection of the KiDS images, is dubbed ''Lenses in KiDS" (LinKS). KiDS is a particularly suitable survey for finding strong lenses, given its excellent seeing and pixel scale, in addition to the large sky coverage (see \Sec\ref{sec:KiDS}).

The paper is organized as follows. In \Sec\ref{sec:KiDS} we provide a brief description of the KiDS survey and the way in which we select the LRG-galaxy sample used in this work. In \Sec\ref{SECtraining} we illustrate our lens-finding CNN-based algorithm and how we build the training data set. In \Sec\ref{SECresults} we explain how we apply our method to $\sim255$ square degrees of KiDS, present the list of our new lens candidates, compare it with the literature and with a forecast of the expected number of detectable strong gravitational lenses in the survey and do a consistency check of the observed Einstein radii of the candidates to select the most reliable ones. Finally, in \Sec\ref{SECconclusions}, we provide a summary, the main conclusion of this work and a short outlook for future plans and improvements.
In the following we adopt a cosmological model with $(\Omega_{m},\Omega_{\Lambda},h)=(0.3,0.7,0.75)$, where $h = H_{0}/100 \, \textrm{km} \, \textrm{s}^{-1} \, \textrm{Mpc}^{-1}$. 

\section{The KiDS survey}\label{sec:KiDS}
The Kilo-Degree Survey (KiDS) \citep{deJong+15_KiDS_paperI} is one of the three ESO public surveys carried out using the OmegaCAM wide-field imager 
(\citealt{Kuijken11}) 
mounted at the Cassegrain focus of the VLT Survey Telescope (VST; \citealt{Capaccioli_Schipani11}) 
at Paranal Observatory in Chile. OmegaCAM is a 256 Megapixel camera containing 32 science CCD detectors which cover a one square degree field of view at a pixel-size of  0.21 arcsec. The VST is a 2.6m telescope with active control of the primary and secondary mirror which is driven by wave-front sensing via two auxiliary CCDs in OmegaCAM. In this way, the camera-telescope combination is specifically designed to obtain sharp and homogeneous image quality over the wide field of view. KiDS is a 1500 square degree extra-galactic imaging survey in four optical bands ($u$, $g$, $r$ and $i$). The survey area is divided over an equatorial patch and a Southern patch around the South Galactic Pole. Observations are queue scheduled, reserving the best seeing for the $r$-band which has a median FWHM of the PSF of 0.65 arcsec with a maximum of 0.8 arcsec. Median PSF FWHM values in $u$, $g$ and $i$ are 1.0 arcsec, 0.8 arcsec and 0.85 arcsec, respectively. KiDS reaches limiting magnitudes (5-$\sigma$ AB in a 2 arcsec aperture) of 24.3, 25.1, 24.9 and 23.8 in $u$, $g$, $r$ and $i$ band, respectively. The primary science driver for the survey design is the study of the dark matter distribution over cosmological volumes via weak-lensing tomography. Strong-lensing survey studies are a particularly suitable science case as well, because they exploit the combination of superb image quality and wide survey area. 

\begin{figure}
  \begin{center}
  \centering
  {\includegraphics[width=90mm]{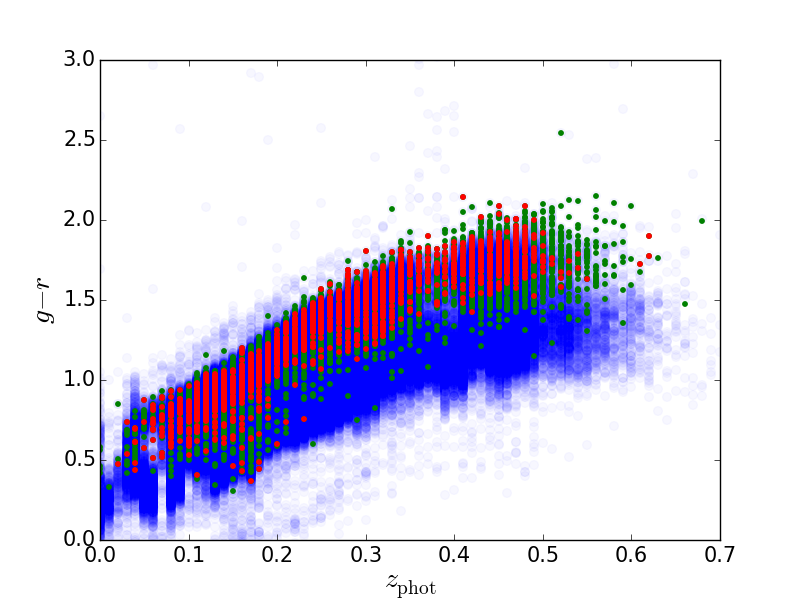}}
\caption{colour $g - r$ versus photometric redshift. The $g$ and $r$ values are \Mauto\ magnitudes and the photometric redshift is obtained with \textsc{BPZ}. The dots are sources from KiDS DR3. Shown are (i) extended objects with \Mauto\ in $r$-band less than 20 (blue), (ii) objects that satisfy the \protect\cite{Eisenstein2001} colour-magnitude selection (red) and (iii) objects selected with our expanded colour-magnitude selection (green). See \Sec\ref{sec:LRG} for the details.
}
\label{FIGcolourz}
\end{center}
\end{figure}

\subsection{Data Release Three}
In this paper we make use of the most recent public data release (KiDS ESO-DR3, de Jong et al. 2016, in prep). It consists of the co-added images, weight maps, masks, single-band and multi-band catalogues and photometric redshifts for 292 survey tiles. We use the multi-band photometry based on $r$-band detections, with a total of 33 million unique sources. Our data handling and scientific data analysis is performed using the Astro-WISE information system \citep{Valentijn2007}. 
The source extraction and related photometry have been obtained with \textsc{S-Extractor} (\citealt{Bertin_Arnouts96_SEx}). We rely on both aperture photometry and the Kron-like {\tt MAG\_AUTO}. A relevant output parameter of \textsc{S-Extractor} is the {\tt FLAGS} parameter. We set the $r$-band {\tt FLAGS} to be $< 4$, to only include de-blended sources and remove from the catalogues those objects with incomplete or corrupted photometry, saturated pixels or any other kind of problem encountered during de-blending or extraction. Critical areas as saturated pixels, star spikes and reflection halos have been masked using a dedicated automatic procedure (\textsc{Pulecenella}). The {\tt IMA\_FLAGS} flags store the result of this masking operation: sources that are not in critical regions have this parameter set to 0. Photometric redshifts are determined using the program \textsc{BPZ} (\citealt{Benitez00}), which is a Bayesian photo-$z$ estimator based on a template fitting method (see de Jong et al. 2017, in prep., for further details). 
The unmasked effective area adopted, considering the sources with {\tt IMA\_FLAGS} $=0$ in all the KiDS-DR3 bands, is 255 square degrees.

\subsection{Luminous red galaxy sample}\label{sec:LRG}
We select Luminous Red Galaxies (LRGs; \citealt{Eisenstein2001}) from the 255 square degrees of the KiDS-ESO DR3 for the purpose of both training our CNN and searching for lens candidates among them. LRGs are very massive and hence more likely to exhibit lensing features compared to other classes of galaxies ($\sim 80\%$ of the lensing population; see \citealt{turner1984statistics,fukugita1992statistical,kochanek1996flat,chae2003cosmic,oguri2006image,moller2007strong}). We focus on this kind of galaxies in this work and will consider other kind of galaxies in the future.
The selection is made with the following criteria where all the parameters are from \textsc{S-Extractor} and magnitudes are \Mauto:\\

\noindent (i) The low-$z$ ($z<0.4$) LRG colour-magnitude selection of \cite{Eisenstein2001}, adapted to including more sources (fainter and bluer): 
\begin{equation}
\begin{split}
&r<20 \\
&|c_{\rm{perp}}| < 0.2 \\
&r<14+c_{\rm{par}}/0.3 \\
\text{where}\\
&c_{\rm{par}}=0.7(g-r)+1.2[(r-i)-0.18)]\\
&c_{\rm{perp}}=(r-i)-(g-r)/4.0-0.18
\end{split}
\end{equation}

\noindent (ii) A source size in the \textit{r}-band larger than the average FWHM of the PSF of the respective tiles, times a empirical factor to maximize the separation between stars and galaxies.

This final selection provides an average of 74 LRGs per tile and a total of 21789 LRGs. We refer to this sample as the "LRG sample" in the remainder of the paper. Compared to the original colour-magnitude selection for $z<0.4$ \citep{Eisenstein2001}, we obtain $\sim 3$ times more galaxies. A colour-photo-$z$ diagram of the results of the two different cuts is shown in \Fig\ref{FIGcolourz} for illustration. 

\section{Training the CNN to find Lenses}\label{SECtraining}

Our lens finder is based on a Convolutional Neural network (CNN) and is inspired by the work of \cite{dieleman2015rotation}. 
CNNs are supervised deep learning algorithms (see the recent reviews from \citealt{schmidhuber2015deep,lecun2015deep,guo2016deep}) particularly effective for image recognition tasks (see e.g., \citealt{he2015deep}, winner of the last ILSVRC competition; \citealt{ILSVRC15}) and regression tasks, such as, in the astronomical domain,
the determination of galaxy morphologies \citep{dieleman2015rotation,Huertas2015}. The algorithm converts sequentially the input data through non-linear transformations whose parameters are learned in the training phase. A set of labelled images (the training set) are used as input of the CNN in this phase. The network changes its parameters by optimizing a loss function that expresses the difference between its output and the labels of the images in the training set. This allows the CNN to learn complex functions and to extract features from the data that are not hand designed but are learned during the training stage. After the training procedure the CNN can be used for classifying new data by keeping its parameters fixed. For the interested reader, in Appendix \ref{SECbackground} we shortly introduce the technical background of CNNs that are relevant to some of the the choices made in this paper.   

\begin{table}
\caption{The range of values adopted for the model parameters of the lens and source. See \Sec\ref{sec:sims} for further details.}
	\begin{center}
		\begin{tabular}{l l c}
Parameter              & Range & Unit \\
\hline
\multicolumn{3}{c}{Lens (SIE)}\\
\hline
Einstein radius      & 1.4 - 5.0 & arcsec\\
Axis ratio           & 0.3 - 1.0  & -\\
Major-axis angle     & 0.0 - 180 & degree\\
External shear       & 0.0 - 0.05 & -\\
External-shear angle & 0.0 - 180 & degree\\
\hline
\multicolumn{3}{c}{Source (S\'ersic)}\\
\hline
Effective radius     & 0.2 - 0.6 & arcsec\\
Axis ratio           & 0.3 - 1.0 & -\\
Major-axis angle     & 0.0 - 180 & degree\\
S\'ersic index       & 0.5 - 5.0 & -\\
\hline
		\end{tabular}
		\label{TABLEmocksourceslens}
	\end{center}
\end{table}

\subsection{Input Samples}

Finding strong gravitational lenses can be reduced to a two-class classification problem, where the two kinds of objects to recognize are the lenses and the non-lenses. Training a Convolutional Neural Network (CNN) to solve this task requires a dataset representative of the two classes called \textit{training set}. It has to be large enough because of the large number of parameters of a CNN (usually of the order of $10^6$). In the case of strong gravitational lenses we do not have a large enough representative data-set at our disposal. The largest sample available is collected in The Masterlens Database\footnote{\href{http://masterlens.astro.utah.edu/}{\tt http://masterlens.astro.utah.edu/}}. Unfortunately, this sample can not be used as a training set for our purpose, since it is small and heterogeneous. It consists of 657 lens systems that are not all spectroscopically confirmed, that have been discovered in various surveys and programs, or that are observed at different wavelengths according to the instrument used.

\begin{figure}
\begin{center}
\centering
\hspace{-1cm}
\includegraphics[width=90mm]{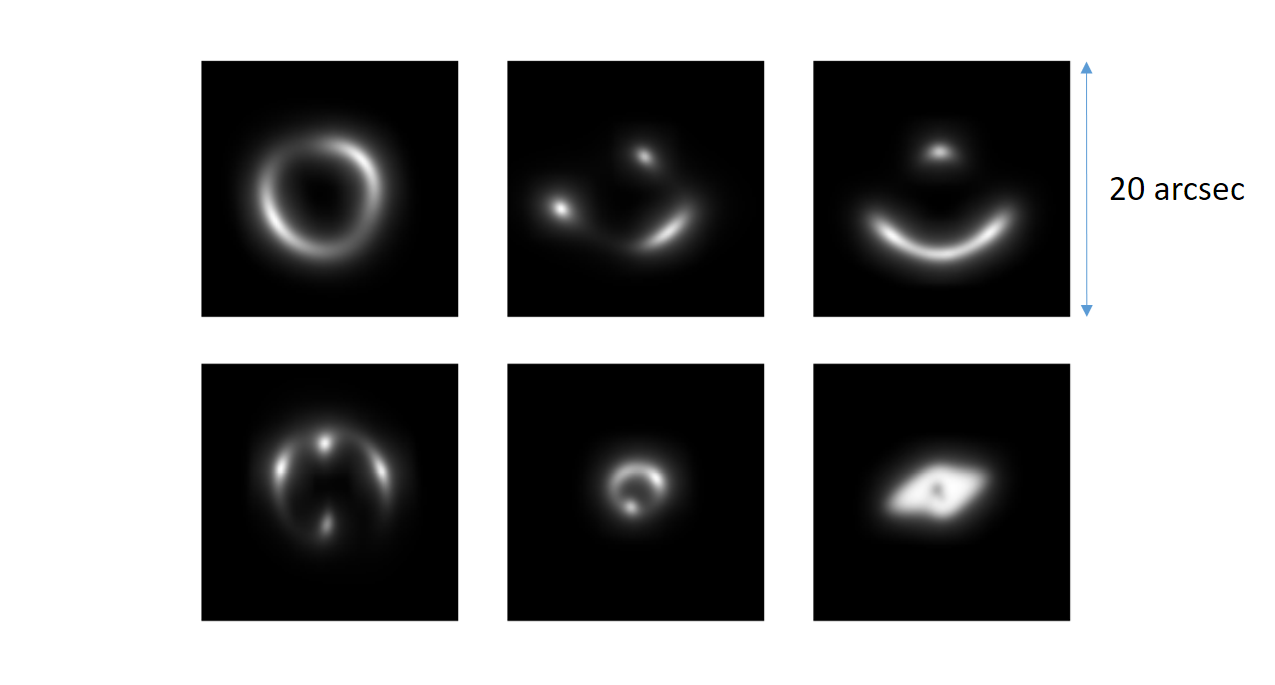}
\caption{Several examples of simulated lensed sources produced as described in \Sec\ref{sec:sims}. The image size is 101 by 101 pixels, corresponding to 20 by 20 arcsec.}
\label{FIGsimulatedsources}
\end{center}
\end{figure}

For these reasons, we build a set of mock lens systems, relying on a hybrid approach: first we select real galaxies, with their fields, obtained from KiDS (\Sec\ref{sec:realsample}), in order to include seeing, noise and especially the lens environment that is a feature hard to simulate and its omission would limit the ability of the network to recognize lenses in real survey data. Then we independently simulate the lensed sources (\Sec\ref{sec:sims}) and combine them with the real galaxies (\Sec\ref{sec:resume}).

We limit our training to $r$-band images, where KiDS provides the best image quality (an average FWHM of 0.65 arcsec). Hence, the network will learn selection criteria mostly based on the morphology of the sources. We plan to ingest multi-wavelength data into the network in future improvements, allowing the training on the differences in colours. Our training set consists of images of lens and non-lens examples produced with \textit{r}-band KiDS images of real galaxies (see \Sec\ref{sec:realsample}) and mock gravitational lensed sources (see \Sec\ref{sec:sims}). 
In \Sec\ref{sec:resume} we summarize how the actual positive (lenses) and negative examples (non-lenses) employed in the training of the network, are produced. We train our CNN on a set of six millions images (three million lenses and three million non-lenses with labels 1 and 0, respectively). Our trained CNN gives as output a value $p$ ranging between 0 and 1. The sources with an output value of $p$ larger than 0.5 are classified as lenses. The technical details of our implementation and the training procedure can be found in Appendix \ref{SEC:appendix}, providing further background to our procedures and choices.
We further expand our training set using data augmentation techniques (\Sec\ref{sec:Dataugm}). 

\subsubsection{Real Galaxy Sample}\label{sec:realsample}
We select a sub-sample of the KiDS LRGs (see \Sec\ref{sec:LRG}) consisting of 6554 galaxies (a third of the full sample), which we have visually inspected finding 218 contaminants, mostly face-on spirals.
Additionally, we have collected a sample of 990 sources wrongly classified as lenses in previous tests with CNNs. We use this sample in the training set to reject clear outliers.
The 6326 LRGs, the 218 contaminants and the 990 false positives constitute together the non-simulated part of the data used to build the training set. We will refer to it as the real galaxy sample in the remaining of this paper.

\subsubsection{Mock Lensed-Source Sample}\label{sec:sims}
The mock lensed source sample is composed by $10^6$ simulated lensed images of 101 by 101 pixels, using the same spatial resolution of KiDS (0.21 arcsec per pixel), corresponding to a 20 by 20 arcsec field of view. We produce the different lensed image configurations by sampling uniformly the parameters of the lens and source models listed in \Tab\ref{TABLEmocksourceslens}. A few examples are shown in \Fig\ref{FIGsimulatedsources}.
The choice of uniformly sampling the parameter space does not reproduce the distribution of the parameters for a real lens population, but allows the classifier to learn the features for recognizing the different kinds of lenses, no matter how likely they are to appear in a real sample of lenses.

We model the sources with a \cite{Sersic68} profile and the lenses with a Singular Isothermal Ellipsoid (SIE; \citealt{KSB_SIE94}) model. At source redshifts of $z>0.5$, smaller sizes and smaller S\'ersic indices are found with respect to the local universe, and the fraction of spiral galaxies (with $n <2-3$) increases (e.g. \citealt{Trujillo+07}; \citealt{Chevance+12}). We exclude spiral galaxy sources or very elliptical ones considering only axis ratios $> 0.3$. The source positions are chosen uniformly within the radial distance of the tangential caustics plus one effective radius of the source S\'ersic profile. This leads our training set to be mostly composed of high-magnification rings, arcs, quads, folds and cusps rather than doubles \citep{schneider1992gravitational} that are harder to distinguish from companion galaxies and other environmental effects. In this paper our first-order goal is to find the larger, brighter and more magnified strong lenses, rather than aim for completeness over the full parameter space of lenses.

The upper limit of 5 arcsec for the Einstein radius aims to include typical Einstein radii for strong galaxy-galaxy and group-galaxy lenses (\citealt{koopmans2009}; \citealt{Foex+13_SARCS}; \citealt{Verdugo+14_SL2S}). The lower limit is chosen to be 1.4 arcsec, about twice the average FWHM of the \textit{r}-band KiDS PSF. Because lenses are typically early-type galaxies, which do not have high ellipticity, we choose 0.3 as a lower limit of the axis ratio (\citealt{Binney_Merrifield98_book}). We set the external shear to less than 0.05, higher than typically  found for SLACS lenses (\citealt{koopmans2006}) with a random orientation varying between 0 and 180 degrees.

\subsection{Building the training examples}\label{sec:resume}
\begin{figure}
\begin{center}
\centering
\includegraphics[width=90mm]{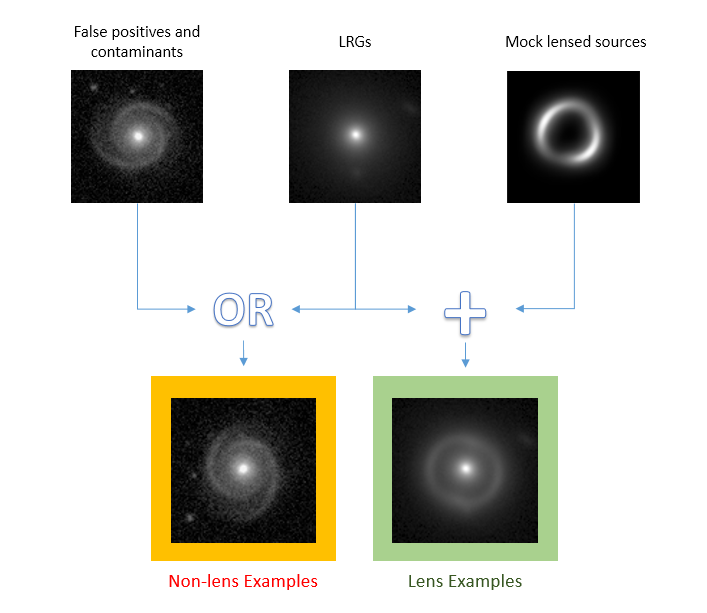}
\caption{A schematic of the training-set creation. For the non-lens examples we use real KiDS image-cutouts of LRGs and other galaxies (see \Sec\ref{sec:realsample}). For producing the lens examples we mix KiDS LRGs and simulated mock lensed sources (\Sec\ref{sec:sims}). In the process the images are augmented and preprocessed as explained in \Secs\ref{sec:Dataugm} and \ref{sec:resume}.}\label{Figdiagram}
\end{center}
\end{figure}

Each training image passed to the network is built as described below and as summarized schematically in \Fig\ref{Figdiagram}.\\

\noindent {\bf Mock lenses (positive sample)}:
To create the mock lenses we carry out the following procedure: (i) we randomly choose a mock lensed source from the mock source sample and a LRG from the real galaxy sample (\Secs\ref{sec:sims} and \ref{sec:realsample}, respectively); (ii) we randomly perturb both the mock source and the LRG as described in \Sec\ref{sec:Dataugm}; (iii) we rescale the peak brightness of the simulated source between 2\% and 20\% of the peak brightness of the LRG. In this way we take into account the typical lower magnitudes of the lensing features with respect to the lens galaxies despite the magnification; (iv) we add the two resulting images; (v) we clip the negative values of the pixels to zero and performing a square-root stretch of the image to emphasize lower luminosity features; and (vi) finally we normalize the resulting image by the peak brightness.
This procedure can yield a-typical lens configurations, because the mock sources and the KiDS galaxies are combined randomly, without taking into account the physical characteristics of the galaxies. Nevertheless, we operate in this way with the intent to train the network to classify a lens largely relying on the morphology of the source. Moreover, we reduce the risk of over-fitting, because the probability that the network will see twice the same (or a very similar) example is negligible. In addition, we cover the parameter space as free from priors as possible, which could allow to find less conventional lens configurations as well.\\

\captionsetup[subfigure]{labelformat=empty}
 \begin{figure*}
   \centering
   \hspace{\fill}
   \subfloat[]{\includegraphics[width=28mm]{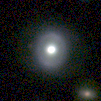}}\hspace{\fill}
   \subfloat[]{\includegraphics[width=28mm]{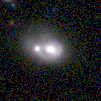}}\hspace{\fill}
   \subfloat[]{\includegraphics[width=28mm]{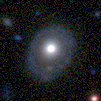}}\hspace{\fill}
   \subfloat[]{\includegraphics[width=28mm]{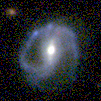}}\hspace{\fill}
   \subfloat[]{\includegraphics[width=28mm]{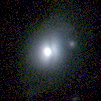}}\hspace{\fill}
\caption{RGB images of 20 by 20 arcsec of some contaminants classified as lenses by the CNN.}
 \label{FIGcontaminants}
 \end{figure*}

\noindent \textbf{Non-lenses (negative sample)}: To create the mock non-lens sample we carry out the following procedure: (i) we randomly choose one galaxy from the real galaxy sample (see \Sec\ref{sec:realsample}) with a 60\% probability of extracting a LRG and 40\% probability to extract a contaminant or false positive; (ii) we randomly perturbing as in \Sec\ref{sec:Dataugm}; (iii) we apply a square-root stretch of the image; (iv) we normalizing the image by the peak brightness.\\

The final inputs of the convolutional neural network are image-cutouts of 60 by 60 pixels which correspond to $\sim 12$ by 12 arcsec. These images are produced in real-time during the training phase.

\subsection{Data augmentation}\label{sec:Dataugm}
A common practice in machine learning is data augmentation: a procedure used to expand the training set in order to avoid over-fitting the data and teaching the network rotational, translational and scaling invariance (see e.g., \citealt{Simard2003}). We augment our dataset applying the following transformations to the mock lensed images and the real galaxy sample: (i) a random rotation between 0 and $2\pi$; (ii) a random shift in both $x$ and $y$ direction between -4 and +4 pixels; (iii) a $50\%$ probability of horizontally flipping the image; (iv) a rescaling with a scale factor sampled log-uniformly between $1/1.1$ and $1.1$.
All transformations are applied to the image-cutouts of 101 by 101 pixels of both the real galaxy and mock lensed source sample. We extract a central region of 60 by 60 pixels from the resulting images to avoid unnecessarily information (i.e., noise and empty sky) around the image edges.

\section{Results}\label{SECresults}

Having trained the CNN as described in Sect.~\ref{SECtraining} (see also Appendix~\ref{SEC:appendix} for more details),
in this section we present our results. In \Sec\ref{SECvisual} we report the procedure to select our final sample of lens candidates and in \Sec\ref{SECradiuscheck} the sample is presented, discussed and compared with the literature.

\subsection{Candidate selection}\label{SECvisual}

First we ingest the full 21789 LRG sample (see \Sec\ref{sec:LRG}) in to the trained CNN. We obtain 761 galaxies ($\sim3\%$ of the full LRG sample) classified as lens candidate with $p>0.5$ and all the remainder in the non-lens category with  with $p<0.5$. The number of LRG classified by the network as lenses is too large when compared to the expected number of strong lenses in the KiDS-DR3 area (see \Sec\ref{sec:expected}). Among the selected sources there are contaminants such as spirals, galaxies with dust lanes, mergers, etc. (see \Fig\ref{FIGcontaminants} for some examples). For this reason we decide to further visually classify the 761 targets selected by the network. Seven of the authors of this paper -- referred as ``classifiers" in the following -- are presented with a set of images for each lens candidate: the cut-out images from KiDS (one image per each of the \textit{u}, \textit{g}, \textit{r}, and \textit{i} filters) and a RGB reconstructed composite image obtained with the software STIFF\footnote{\href{http://www.astromatic.net/software/stiff}{\tt http://www.astromatic.net/software/stiff}} from the \textit{g}, \textit{r}, and \textit{i}-band images.
The classifiers can classify the sources in three categories: \textit{Sure}, \textit{Maybe}, and \textit{No lens}. The score for each candidate is based on the following scheme:\\

\begin{tabular}{ll}
\textit{Sure lens} & 10 points. \\
\textit{Maybe lens}  & 4 points. \\
\textit{No lens} &  0 points. \\
\end{tabular}\\

\begin{figure}
  \begin{center}
  {\includegraphics[width=80mm]{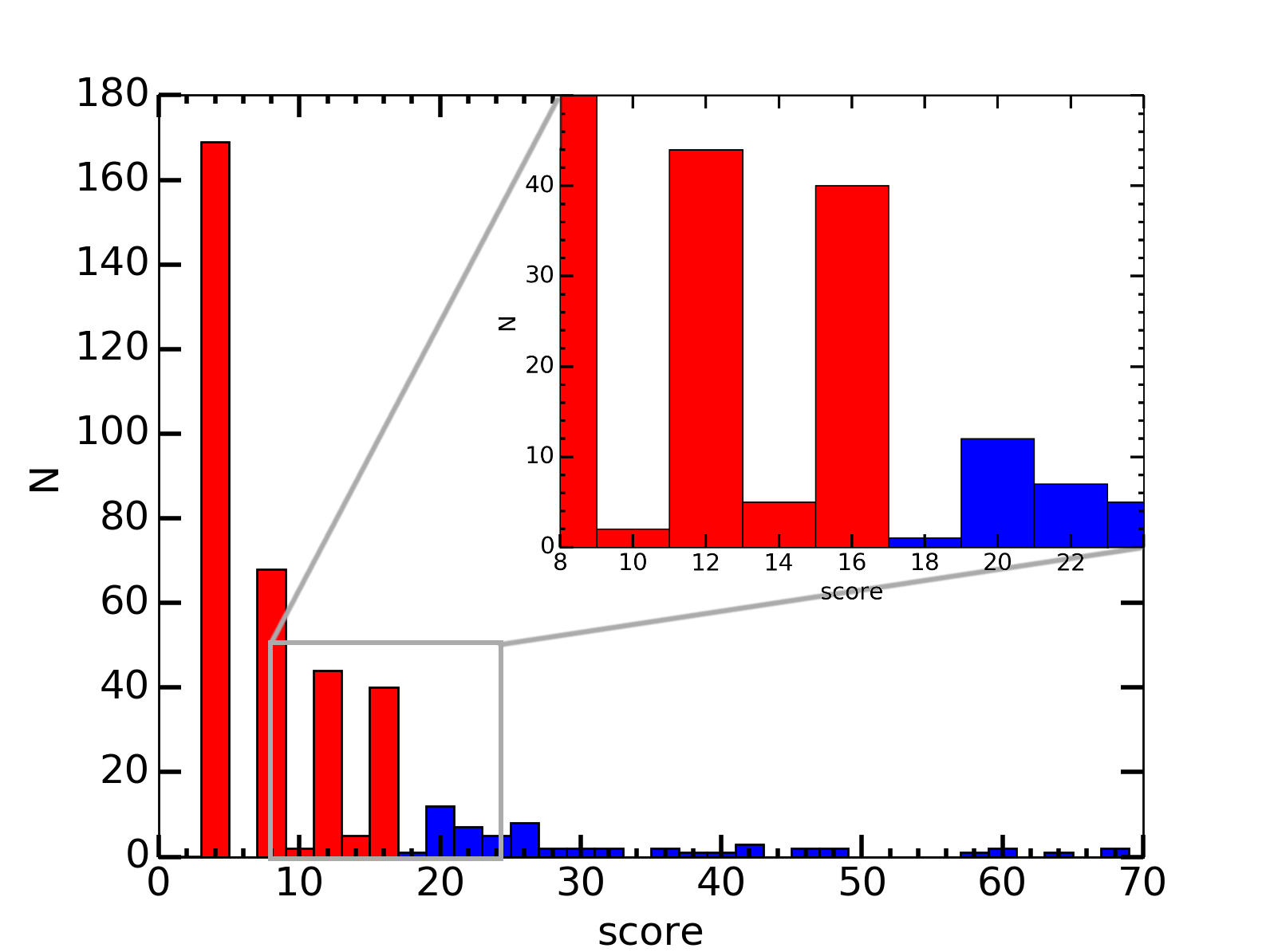}}
\caption{Histogram of the ranking of 384 lens candidates, which have been classified at least by one user in the \textit{Sure} or \textit{Maybe} categories. In blue are the candidates with a score higher than 16 that are considered the most reliable.}
\label{FIGrankmerged}
\end{center}
\end{figure}

\noindent The histogram of the accumulated grades of the visual classification is shown in \Fig\ref{FIGrankmerged}. There are 384 candidates classified in the \textit{Sure} and \textit{Maybe} categories by at least one classifier. To further reduce the sample, we decide to introduce a threshold at the score of 17, below which all candidates are considered not reliable. This implies that more than four classifiers would be required to classify a lens candidate in the \textit{Maybe} category to be regarded as reliable. For lenses in the \textit{Sure} category we expect a large number of users to agree in their classification due to more evident lensing features in the images, giving a higher score to such candidates. Only two candidates achieved the maximum score of 70. As seen in \Fig\ref{FIGrankmerged} (blue bars), the distribution of candidates rises rapidly below the threshold score and remains flat for higher values. Changing the points given to a candidate classified as \textit{Maybe} lens from four to six, and appropriately relocating the threshold, does not affect the resulting ranking, and the distribution shown in \Fig\ref{FIGrankmerged} remains largely the same.

Since the focus of this paper is to find new lens candidates, we are interested in the first two categories, i.e. \textit{Sure} and \textit{Maybe}. However, we plan for future applications to use the candidates classified in the \textit{No} category to retrain the CNN, aiming at considerably reducing the number of candidates that need to be visually inspected.

\subsection{Final sample of candidates}\label{SECradiuscheck}

\begin{figure}
  \begin{center}
  {\includegraphics[width=80mm]{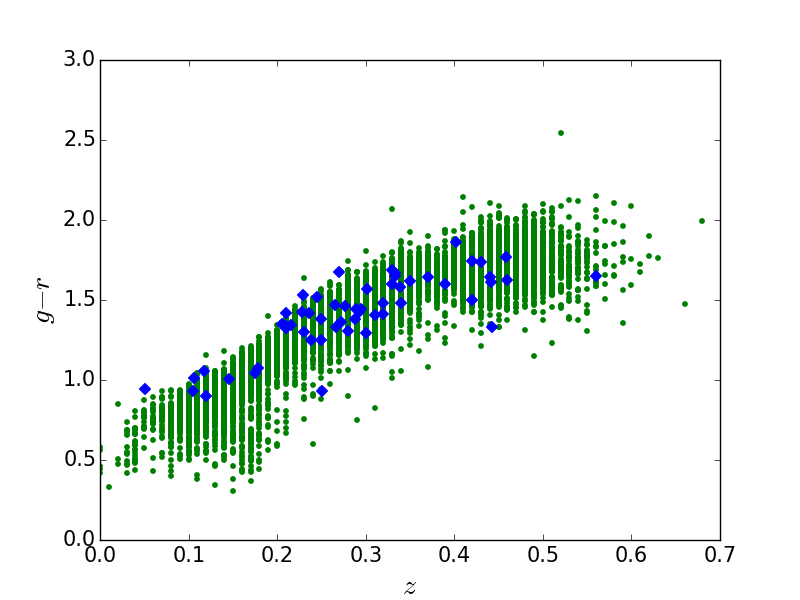}}
\caption{\textit{g-r} colour-redshift distribution of the LRG sample (green dots; \Sec\ref{sec:LRG}) and our 56 best candidates (blue diamonds; \Sec\ref{SECresults}). The BPZ photometric redshift is plotted, except for the candidates with an available spectroscopic redshift.}
\label{FIGcolredcand}
\end{center}
\end{figure}

After both CNN and visual classification, the final sample of lens candidates consists of 56  objects, down-selected from an initial sample of 21789 galaxies. In \Fig\ref{FIGcolredcand} we show how the candidates are distributed in colour-photo-$z$ space together with the full LRG sample (\Sec\ref{sec:LRG}). In \Fig\ref{FIGCandidatecutouts} the RGB images of these best candidates are shown together with their scores from the visual inspection procedure. For completeness, in Appendix \ref{SEC:appendix2} the \textit{r}-band-only images of the 56 ranked objects are also shown, since they are the images on which the CNN has made its classification. Candidates are listed in \Tab\ref{TABLEcandidates}, where we show the final grade of our classification, the KiDS \Mauto\ in the \textit{u}, \textit{g}, \textit{r}, and \textit{i} bands for each candidate, together with the BPZ photometric redshift, stellar mass and, if available, spectroscopic redshift and velocity dispersion. 

\begin{figure*}
   \centering
   \hspace{\fill}
   \subfloat[J085446-012137]{\includegraphics[width=38mm]{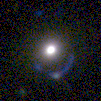}}\hspace{\fill}
   \subfloat[J114330-014427]{\includegraphics[width=38mm]{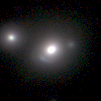}}\hspace{\fill}
   \subfloat[J1403+0006]{\includegraphics[width=38mm]{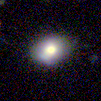}}\hspace{\fill}
\caption{RGB images of the three known lenses present in the LRG sample. The network correctly classifies the first two as lenses, but classifies the third as non-lens. Given that its Einstein radius is 0.83 arcsec, smaller than the Einstein radii of the simulated lenses on which the network has been trained, this might be expected. The images are 20 by 20 arcsec.}
 \label{FIGknownlenses}
 \end{figure*}

J085446-012137 and J114330-014427 are successfully classified as lenses by our network and they pass our visual inspection with a score of 70 and 60 respectively (KSL317 and KSL040 in \Tab\ref{TABLEcandidates} and \Fig\ref{FIGCandidatecutouts}). Instead, J1403+0006 is classified as a non-lens by the network, this could be due to the fact that this system has an Einstein radius of 0.83 arcsec, well below the lower limit of the interval of radii on which the  CNN is trained. In \Fig\ref{FIGknownlenses} we show the RGB images of these three known lenses as observed in KiDS. The lensed images of the misclassified lens are also not as prominent as in the other two. 

We find that 34 of our candidates have spectra measured from different sources (2dF, \citealt{2dfColless2001}; \citealt{Limousin2010}; SDSS, \citealt{SDSS}; BOSS, \citealt{BOSS}; GAMA, \citealt{GAMALiske2015}). 
We visually inspected the spectra without clearly identifying any emission line that could belong to a background source. A more detailed data reduction of the spectra is needed to confirm or discard any of these candidates. We also notice that the photometric redshifts tend to overestimate the distance. This could be due to the contamination of the colours of the main galaxy by the supposed lensed sources. We will investigate this issue in a forthcoming paper.

\begin{table*}
\caption{The final sample of candidates. For each candidate we report an internal ID; the spectroscopic redshift if available (see notes), the BPZ photometric redshift, the KiDS \textit{u}, \textit{g}, \textit{r} and \textit{i} \Mauto\ (average uncertainties are 0.13, 0.02, 0.03 and 0.04 respectively), the velocity dispersion from SDSS or BOSS if available, the stellar mass (the typical uncertainty is $\sim 0.2$ dex). A double (single) check-mark indicates the candidates with (without) a measured velocity dispersion that have a predicted Einstein radius comparable with their galaxy-image configuration (see \Sec\ref{SECradiuscheck}).}
	\begin{center}
		\begin{tabular}{c c c c c c c l c c l}
        ID & $z_{\rm{spec}}$ & $z_{\rm{phot}}$ & $u$  & $g$ & $r$ & $i$ &  \sigs\ [km/s]  & $\log \mst/\Msun$ [dex]  & score &\\
\hline
KSL427 & $0.24^\textit{2}$ & 0.25 & 20.17 & 18.38 & 16.96 & 16.58   &                     & 11.3 & 70 & \checkmark  \\  
KSL317 & $0.35^\textit{4}$ & 0.42 & 21.43 & 19.61 & 17.86 & 17.20   &                     & 11.6 & 70  \\  
KSL103 & $0.24^\textit{2}$ & 0.26 & 20.55 & 18.53 & 17.28 & 16.81   &                     & 11.3 & 64  \\
KSL040 & $0.11^\textit{1}$ & 0.15 & 18.56 & 16.65 & 15.64 & 15.26   &  $269 \pm 5$        & 11.1 & 60 & \checkmark\checkmark \\   
KSL627 & $0.21^\textit{1}$ & 0.24 & 20.89 & 18.75 & 17.42 & 16.92   & $206 \pm 13$        & 11.3 & 60  \\
KSL327 & $0.12^\textit{2}$ & 0.17 & 18.96 & 16.80 & 15.75 & 15.33   &                     & 11.4 & 58 & \checkmark\\
KSL376 & $0.30^\textit{1}$ & 0.36 & 21.88 & 20.03 & 18.46 & 17.91   & $242 \pm 20 $       & 11.2 & 48  \\
KSL086 &                   & 0.33 & 22.30 & 20.23 & 18.54 & 18.03   &                     & 11.1 & 48  \\
KSL351 & $0.26^\textit{1}$ & 0.30 & 21.12 & 18.89 & 17.42 & 16.91   & $278 \pm 19 $       & 11.4 & 46 & \checkmark\checkmark \\
KSL469 & $0.29^\textit{1}$ & 0.33 & 21.23 & 19.52 & 18.08 & 17.51   & $228 \pm 19 $       & 11.4 & 46  \\
KSL228 & $0.18^\textit{2}$ & 0.16 & 19.83 & 18.30 & 17.25 & 16.75   &                     & 11.2 & 42  \\
KSL713 & $0.23^\textit{2}$ & 0.29 & 20.46 & 18.54 & 17.01 & 16.47   & $304 \pm 17 $       & 11.5 & 42  & \checkmark\checkmark \\ 
KSL328 & $0.23^\textit{1}$ & 0.24 & 21.28 & 19.52 & 18.09 & 17.58   & $235 \pm 13 $       & 11.0 & 42  \\ 
KSL411 & $0.25^\textit{1}$ & 0.27 & 18.62 & 17.59 & 16.66 & 16.22   &                     & 11.5 & 40 & \checkmark \\
KSL070 & $0.44^\textit{1}$ & 0.45 & 21.55 & 20.37 & 19.05 & 18.37   & $206 \pm 37  $      & 11.2 & 40  \\
KSL543 &                   & 0.25 & 20.73 & 19.03 & 17.78 & 17.29   &                     & 11.3 & 38  \\
KSL664 &                   & 0.30 & 21.89 & 19.91 & 18.62 & 18.00   &                     & 11.1 & 36  & \checkmark  \\
KSL106 & $0.27^\textit{3}$ & 0.28 & 21.75 & 19.96 & 18.59 & 18.06   &                     & 11.1 & 36  & \\ 
KSL337 &                   & 0.35 & 22.06 & 20.55 & 18.93 & 18.41   &                     & 10.8 & 32  & \\
KSL388 & $0.33^\textit{1}$ & 0.37 & 22.83 & 19.81 & 18.14 & 17.58   & $228 \pm 20 $       & 11.4 & 32  & \\
KSL415 & $0.21^\textit{1}$ & 0.21 & 20.64 & 19.03 & 17.68 & 17.19   & $223 \pm 17 $       & 11.3 & 32  & \\ 
KSL220 &                   & 0.31 & 21.85 & 20.32 & 18.91 & 18.42   &                     & 11.2 & 30  & \checkmark \\
KSL601 & $0.46^\textit{1}$ & 0.54 & 23.09 & 21.41 & 19.65 & 18.97   &  $221 \pm 22 $      & 11.1 & 28  & \\ 
KSL603 & $0.34^\textit{1}$ & 0.41 & 21.96 & 20.04 & 18.46 & 17.86   &  $220 \pm 16 $      & 11.5 & 28  & \\
KSL436 &                   & 0.27 & 21.44 & 19.57 & 17.90 & 17.74   &                     & 10.8 & 26  & \\
KSL233 & $0.15^\textit{2}$ & 0.17 & 19.96 & 18.17 & 17.16 & 16.67   &                     & 10.8 & 26  & \\
KSL231 &                   & 0.46 & 23.73 & 20.98 & 19.36 & 18.64   &                     & 11.3 & 26  & \\
KSL101 &                   & 0.32 & 22.92 & 20.36 & 18.89 & 18.33   &                     & 11.2 & 26  & \checkmark \\
KSL450 & $0.40^\textit{1}$ & 0.46 & 23.24 & 20.83 & 18.97 & 18.44   &  $270 \pm 30 $      & 11.2 & 26  & \checkmark\checkmark \\
KSL737 & $0.37^\textit{1}$ & 0.45 & 22.27 & 20.39 & 18.74 & 18.14   &  $222 \pm 38 $      & 11.4 & 26  & \checkmark\checkmark \\ 
KSL094 & $0.29^\textit{1}$ & 0.42 & 20.77 & 19.36 & 17.94 & 17.37   &  $219 \pm 19 $      & 11.5 & 26  & \checkmark\checkmark \\
KSL669 & $0.05^\textit{1}$ & 0.16 & 19.46 & 17.58 & 16.63 & 16.17   &   $212 \pm 8  $     & 10.4 & 26  & \\
KSL707 &                   & 0.25 & 21.77 & 19.66 & 18.28 & 17.72   &                     & 11.0 & 24  & \\
KSL197 &                   & 0.21 & 21.63 & 19.56 & 18.14 & 17.71   &                     & 11.1 & 24  & \\
KSL335 & $0.22^\textit{2}$ & 0.26 & 21.89 & 19.07 & 17.72 & 17.19   &                     & 11.2 & 24  & \\
KSL565 & $0.29^\textit{1}$ & 0.29 & 21.52 & 19.97 & 18.52 & 17.97   &   $251 \pm 18 $     & 11.1 & 24  & \checkmark\checkmark \\   
KSL134 & $0.27^\textit{1}$ & 0.29 & 21.23 & 19.57 & 18.24 & 17.73   &   $235 \pm 14  $    & 11.3 & 24  & \\
KSL606 & $0.18^\textit{2}$ & 0.17 & 19.79 & 18.32 & 17.25 & 16.85   &                     & 11.2 & 22  & \\
KSL046 &                   & 0.12 & 18.44 & 16.69 & 15.79 & 15.40   &                     & 11.4 & 22  & \checkmark \\
KSL620 &                   & 0.28 & 21.53 & 19.78 & 18.47 & 17.91   &                     & 11.1 & 22  & \\
KSL013 & $0.11^\textit{2}$ & 0.14 & 19.73 & 18.19 & 17.26 & 16.74   &                     & 10.7 & 22  & \\
KSL421 &                   & 0.32 & 21.42 & 19.65 & 18.23 & 17.69   &                     & 11.4 & 22  & \checkmark \\
KSL434 &                   & 0.39 & 21.74 & 20.06 & 18.47 & 17.99   &                     & 11.0 & 22  & \\
KSL516 &                   & 0.56 & 23.33 & 21.19 & 19.55 & 18.80   &                     & 11.1 & 20  & \\
KSL278 &                   & 0.42 & 21.85 & 20.34 & 18.84 & 18.32   &                     & 11.2 & 20  & \checkmark \\
KSL178 &                   & 0.43 & 21.74 & 20.04 & 18.31 & 17.63   &                     & 11.9 & 20  & \\
KSL159 &                   & 0.44 & 22.08 & 20.32 & 18.68 & 18.08   &                     & 11.7 & 20  & \\
KSL686 & $0.25^\textit{3}$ & 0.30 & 21.77 & 19.24 & 17.72 & 17.20   &                     & 11.2 & 20  & \\
KSL465 &                   & 0.34 & 20.49 & 19.17 & 17.69 & 17.17   &                     & 11.6 & 20  & \checkmark \\
KSL463 &                   & 0.23 & 21.24 & 19.61 & 18.31 & 17.85   &                     & 10.8 & 20  & \checkmark \\
KSL342 &                   & 0.21 & 21.22 & 19.28 & 17.95 & 17.49   &                     & 10.9 & 20  & \checkmark \\
KSL322 & $0.33^\textit{3}$ & 0.44 & 22.27 & 20.02 & 18.37 & 17.76   & $333 \pm 25 $       & 11.5 & 20  & \checkmark \checkmark \\
KSL674 & $0.28^\textit{1}$ & 0.31 & 21.68 & 19.60 & 18.14 & 17.60   &   $293 \pm 21 $     & 11.2 & 20  & \checkmark \checkmark\\
KSL564 & $0.29^\textit{1}$ & 0.33 & 22.92 & 19.85 & 18.47 & 17.88   &   $249 \pm 22 $     & 11.2 & 20  & \checkmark\checkmark \\
KSL670 & $0.44^\textit{1}$ & 0.48 & 23.71 & 21.10 & 19.49 & 18.72   & $207 \pm 25 $       & 11.4 & 20  &  \\
KSL535 &                   & 0.44 & 22.08 & 20.32 & 18.68 & 18.08   &                     & 11.5 & 18  & \\                                         
		\end{tabular}
        \begin{tablenotes}
      \small
      \item $^\textit{1}$ \cite{SDSS} and \cite{BOSS}; $^\textit{2}$ \cite{2dfColless2001}; $^\textit{3}$ \cite{GAMALiske2015}; $^\textit{4}$ \cite{Limousin2010}

    \end{tablenotes}
		\label{TABLEcandidates}
	\end{center}
\end{table*}

\captionsetup[subfigure]{labelformat=empty}

\subsubsection{Expected number of lenses}\label{sec:expected}
To assess whether the amount of selected candidates is reliable, we estimate the detectable lens population in KiDS, using the lens-statistics code \textsc{LensPop}\footnote{\href{https://github.com/tcollett/LensPop}{\tt https://github.com/tcollett/LensPop}} \citep{collet2015}. Assuming an effective KiDS survey area of $\sim1275$ ~sq.~deg we forecast $\sim 2400$ potentially detectable lenses with a total signal-to-noise ratio larger than 20 and having lensed images resolved over least three seeing elements. The expected number of lenses reduces to $\sim 500$ for the effective area of the KiDS-DR3 of  $255$~sq.~deg. If we consider only lenses that satisfy our colour-magnitude cut of \Sec\ref{sec:LRG} and with an Einstein radius $>1.4$ arcsec, i.e., our range of the parameter space, we  forecast $\sim 50$ lenses for $255$~sq.~deg, broadly comparable to the number of our final sample of candidates, especially if we keep in mind that (i) we do not expect the CNN plus human lens selection to be 100\% efficient and (ii) our training was largely focused on arcs and rings and not on quads and doubles. The redshift distribution of our final sample of candidates and the simulated population from \textsc{LensPop}, within the selection constraints specified above, are also consistent. Our candidates are observed in the window $0.1 \lsim z \lsim 0.5$, with a median redshift of $0.28^{+0.12}_{-0.08}$, while the \textsc{LensPop} sample is in the window $0.13 \lsim z \lsim 0.4$ with a median redshift of $0.32^{+0.08}_{-0.09}$. The scatter corresponds to the 16-84th quantiles of the distribution.

\subsection{Sample Characterization and Comparison}

To further characterize the sample of candidates, and allow a comparison with the literature, we estimate the stellar masses using the software \textsc{Le Phare}
(\citealt{Arnouts+99}; \citealt{Ilbert+06}), which performs a $\chi^{2}$ fitting between the stellar population
synthesis (SPS) theoretical models and the data. Single burst models
from \citet[BC03]{BC03} and a \cite{Chabrier01} IMF are implemented in the software. In the BC03 models we leave the age free to vary up to a maximum of $13 \, \rm Gyr$, and assume metallicities in the range (0.005--2.5 \Zsun). No internal extinction is adopted. The single burst models provide us with a fair description of the stellar populations in massive early-type galaxies. Models are redshifted using the photometric redshifts (or the spectroscopic estimates where available). We adopt the observed $ugri$ magnitudes (and related $1\, \sigma$ uncertainties) within
a $5$ arcsec diameter aperture, corrected for Galactic
extinction using the map in \cite{Schlafly_Finkbeiner11}\footnote{These updated extinctions are calculated by multiplying for 0.86 the \cite{SFD98_dust} values stored in the KiDS-DR3 catalog.}. The \textit{r}-band \Mauto\ is used to
correct the outcomes of \textsc{Le Phare} for missing flux. 
For 34 out of the 56 lens candidates (i.e., 60 per cent) we have spectroscopic redshifts. 
As pointed out previously, photometric redshifts tend to be larger than the spectroscopic estimate by $\Delta z \sim 0.04$, on average. 
For the 22 galaxies with a photometric redshift only, an overestimated redshift could imply both an over- or underestimate of the stellar mass, leading to a less reliable stellar mass. We have estimated the average impact of this systematics using the derived masses for the 34 galaxies with both measures of redshifts, finding that the photometric values bring to an average overestimate of the mass of $0.04$ dex, with a scatter of $\sim 0.2$ dex. 
In addition, the aperture photometry adopted for the derivation of stellar masses is also affected. We have considered KiDS magnitudes within a radius of 5 arcsec, thus the enclosed lensing features make bluer colours and thus we underestimate the real stellar mass. However, we expect that this effect is within the typical mass uncertainty, since the arcs are very faint compared to the lens and hence colour contamination is very small. A systematic study of this issue is beyond the scope of this paper: we will discuss its impact on our results in a forthcoming paper.

\begin{figure}
  \begin{center}
  {\includegraphics[width=75mm]{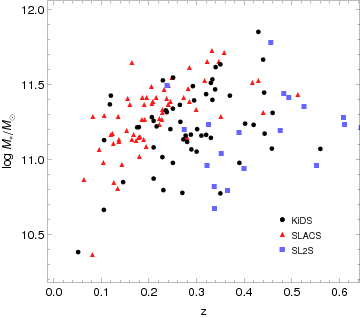}}
\caption{Stellar mass versus redshift for our 56 best candidates (black dots), SLACS sample from \citet[red triangles]{Auger+09_SLACSIX} and SL2S sample from \citet[blue squares]{Sonnenfeld+13_SL2SIV}. The uncertainties on the stellar masses are $\sim 0.1-0.2$ dex.}
\label{FIGmstarvsz}
\end{center}
\end{figure}

In terms of redshift distribution, our lens candidates are observed in the window $0.1 \lsim z \lsim 0.5$, with a median redshift of $0.28^{+0.12}_{-0.08}$. 
This value is larger than the median redshift of SLACS lenses from \cite{Auger+09_SLACSIX}, i.e. $0.20^{+0.09}_{-0.07}$, but consistent within the scatter distribution. Instead, our median redshift is smaller than the average for the SL2S sample from \cite{Sonnenfeld+13_SL2SIV}, i.e. $0.48_{-0.16}^{+0.23}$. This is not surprising given the $z \lsim 0.4$ colour cut of \Sec\ref{sec:LRG}. In future analyses this limit will be loosened.
The median stellar mass of our sample is $\log \mst/\Msun \sim {11.2} \, \rm dex$ with a scatter of $\sim 0.2$ dex. The typical uncertainty of the mass estimates is $\sim 0.1-0.2$ dex too. Within the scatter and mass uncertainties, this value is consistent with the average stellar mass in SLACS (the median is $\log \mst/\Msun \sim  11.3$ dex and the scatter is $\sim 0.2$ dex; \citealt{Auger+09_SLACSIX}) and SL2S lenses (the median is $\log \mst/\Msun \sim  11.2$ dex and the scatter is $\sim 0.25$).  All the galaxies have $\mst \gsim 10^{11} \, \rm \Msun$, except for KSL669 (see next subsection for further comments about this source). 
In \Fig\ref{FIGmstarvsz} we plot the stellar mass as a function of redshift for our sample and the SLACS and SL2S ones.

A similar comparison can be performed for the velocity dispersion, if we consider the KiDS candidates with an available measure of this quantity. The average value for KiDS is $\sigs = 232_{-20}^{+46}\, \rm km/s$\footnote{The velocity dispersions are extracted from both SDSS and BOSS survey, thus, this average value is mixing observations made within two different fibre apertures.}. In SLACS, the average is $\sigs = 243_{-33}^{+47}\, \rm km/s$, while in SL2S the velocity dispersion within a radius of one-half effective radius is $\sigma_{e,2} = 258_{-53}^{+42}\, \rm km/s$. The three estimates agree within the scatter distribution and within the typical uncertainties of velocity dispersion measurements of $\sim 15-20\, \rm km/s$.

\subsubsection{A sanity check of the candidates}\label{SECcheckontheradius}

A sub-sample of candidates has stellar velocity dispersion, $\sigma_{\rm \star}$, measured in the SDSS \citep{SDSS} or BOSS \citep{BOSS} surveys. For these candidates, the knowledge of $\sigma_{\rm \star}$ allows to put constraints on their Einstein radii \RE. For a SIS model, the Einstein radius can be expressed, in radians, as
\begin{equation}
\theta_{\rm E}=4\pi\left(\frac{\sigma_{\rm SIS}}{c}\right)^2\frac{D_{ls}}{D_{s}},
\label{EQeinrad}
\end{equation}
where $D_{s}$, and $D_{ls}$ are, respectively, the angular diameter distances between the observer and the source and between the lens and the source. As a first approximation, in \Eq\eqref{EQeinrad} the $\sigma_{\rm SIS}$ strength parameter can be substituted with the measured stellar velocity dispersion, since they have been found to be approximately equal for lens galaxies (see, e.g., \citealt{bolton2008}). A more rigorous approach consists of deriving the value of $\sigma_{\rm SIS}$ by matching the theoretical velocity dispersion derived from the Jeans equations with the observed one. From the Jeans equations, the radial velocity dispersion can be easily derived. This theoretical quantity is first integrated along the line of sight and then within a circular aperture with SDSS or BOSS fibres radius (i.e., $R_{\rm ap} = 1.5$ and $1$ arcsec, respectively; see \cite{Tortora+09} for equations and further details about the procedure). We use a SIS for the total mass profile, and the light distribution is set adopting a \cite{deVauc48} profile, using the effective radii taken from the SDSS website\footnote{\href{http://skyserver.sdss.org/dr13/en/home.aspx}{\tt http://skyserver.sdss.org/dr13/en/home.aspx}} which come from a de Vaucoulers fit. Imposing that the theoretical aperture-averaged velocity dispersion $\sigma_{\rm Jeans}$ is equal to the observed one $\sigs$, the only free parameter, $\sigSIS$, can be derived. This estimated quantity is finally inserted in \Eq\eqref{EQeinrad}. 

\begin{figure*}
  \begin{center}
  {\includegraphics[width=150mm]{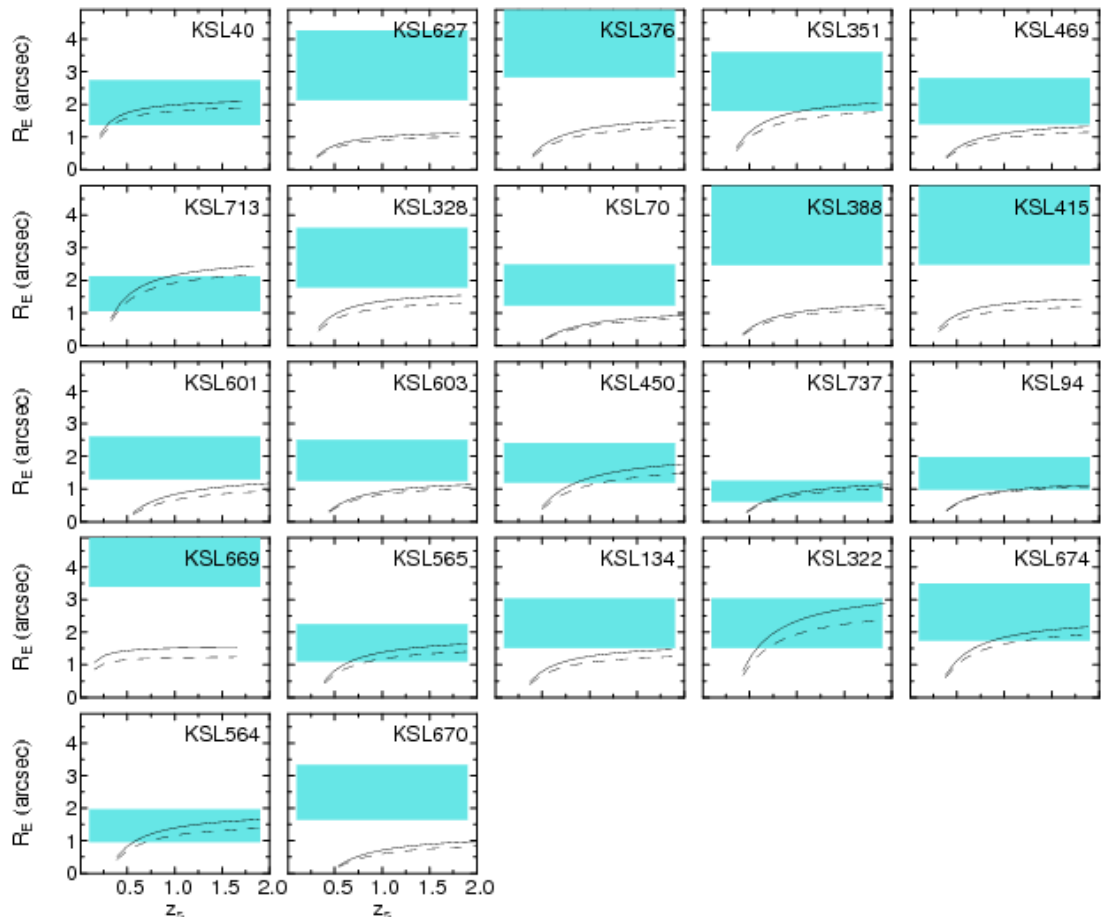}}
\caption{Einstein radii estimated by the observed dynamics are plotted as a function of the unknown source redshift $z_{\rm s}$, and compared with the value estimated visually from the images. The black lines are calculated using \Eq\eqref{EQeinrad}: the solid line assumes that $\sigma_{\rm SIS}$ is determined from Jeans dynamical analysis, and the dashed line by fixing $\sigma_{\rm SIS}=\sigs$ (see \Sec\ref{SECcheckontheradius} for the details). The shaded cyan region corresponds to a conservative range of values for the Einstein radii: it is calculated from the observed distance of the arc from the lens centre, and is set to the range between $0.5R_{\rm E}$ and $R_{\rm E}$.}
\label{FIGeinsteinradii}
\end{center}
\end{figure*}

For each lens with a measured \sigs, the predicted \RE\ is plotted as a function of the unknown $z_{\rm s}$ and compared with the observed Einstein radius (see \Fig\ref{FIGeinsteinradii}). A precise determination of the Einstein radius would require  modelling of the lensing candidates, which is beyond the scope of this paper, but is planned for a follow-up paper. Here, we simply estimate the Einstein radius visually. We take it to be between 1 and 0.5 times the distance between the arc (or the brightest arc in case of multiple images) and the centre of the lens.
This choice is due to the fact that, given a SIS, the image separation for an Einstein ring is exactly twice the Einstein radius, whereas in the case of an arc or an image maximally away from the centre of the lens, the distance from the centre is twice the Einstein radius \citep{KSB_SIE94}. 

In \Fig\ref{FIGeinsteinradii} we show the comparison and we find an overlap for about half of the candidates with measured velocity dispersion. The other half is more likely to be constituted by ring galaxies, foreground sources or other contaminants. An excellent agreement is found for the systems KSL713, KSL450, KSL737, KSL565, KSL322, KSL674 and KSL564. For the other systems, the dynamics predict too small Einstein radii. One interesting case is KSL669, which is a $z=0.05$ galaxy with a very small stellar mass of $\sim 2 \times 10^{10}\, \rm \Msun$. We expect the probability for it to act as a lens to be very small. The comparison performed in \Fig\ref{FIGeinsteinradii} seems to confirm the peculiarity of this lensing candidate. In fact, a larger view of the source shows that it is actually a merger event. 

Moreover, among the best ranked systems, KSL627 and KSL376 present a discrepancy which seems difficult to reconcile. These two systems have almost circular blue rings, with $\sim 4.3$ and $\sim 5.7$ arcsec radii, corresponding to $15$ and $25$ kpc, respectively. These sources do not match typical Einstein radii observed in galaxy-scale gravitational lenses; they are more likely to belong to the category of ring galaxies \citep{Hoag1950,Theys1976,Whitmore1990,bournaud2003,Iodice2003,Madore2009}.

\begin{figure*}
  \begin{center}
  {\includegraphics[width=150mm]{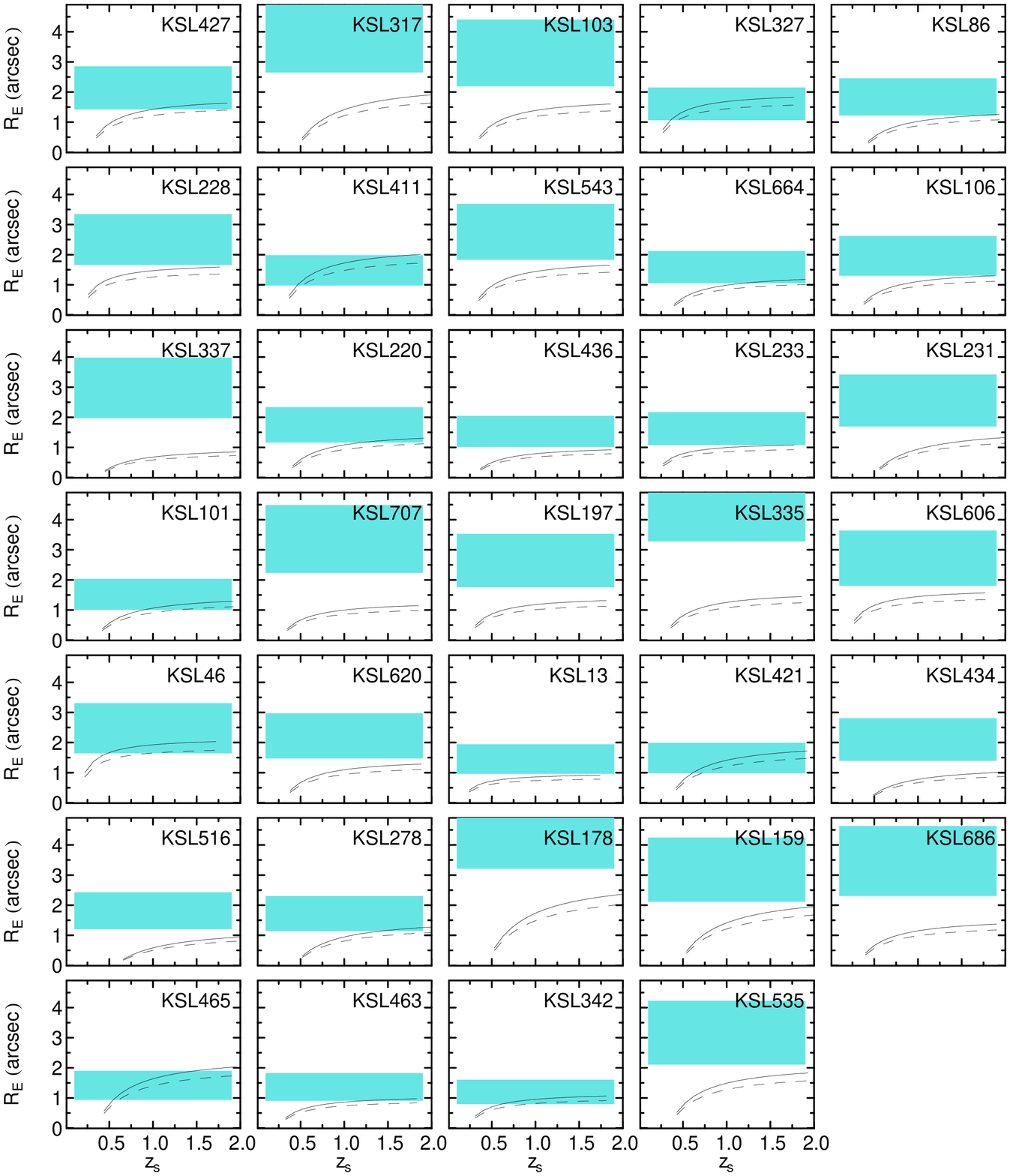}}
\caption{Same as in \Fig\ref{FIGeinsteinradii}, but for the candidates without measured velocity dispersion. In these cases the velocity dispersion is inferred from the stellar mass as described in \Sec\ref{SECcheckontheradius}.}
\label{FIGeinsteinradii2}
\end{center}
\end{figure*}

ETGs follow a tight relationship between velocity dispersion and stellar mass. This can allow us to predict the velocity dispersion for the remaining galaxies in our final sample. After collecting ETG lenses from SLACS \citep{Auger+09_SLACSIX}, we perform a median fit determining the best-fitted relation $\log \sigs = -0.1 + 0.22 \log \mst/\Msun$ between the velocity dispersion and the estimated stellar masses. Thus, assuming that this relation holds for our sample and that it has no scatter, we can determine an estimate for the velocity dispersion, when our stellar masses are used. We use \Eq\eqref{EQeinrad} to predict the Einstein radius as a function of the source redshift. The results are shown in \Fig\ref{FIGeinsteinradii2}, where the dashed line is calculated by inserting the estimated velocity dispersion in \Eq\eqref{EQeinrad}, while the solid line is calculated by assuming the average of the ratios $\sigma_{\rm SIS}/\sigs$ obtained for the galaxies with available velocity dispersion, and inserting the derived $\sigma_{\rm SIS}$ in \Eq\eqref{EQeinrad}. 
Similar considerations as for the galaxies with measured velocity dispersion can be done for these objects, even if the uncertainties on the estimated velocity dispersions are higher.

The previous analysis can give us an indication on the nature of the candidates. However a spectroscopic validation is needed, because it can not be excluded that the lens candidates are part of a group of galaxies. In this case the stellar velocity dispersion would not trace the dynamics of the group. Indeed, this is the case of the known lens J085446-012137 (KSL317), which is part of a group \citep{Limousin2010}, resulting in a under-estimation of the Einstein radius.

We plan to follow-up our most reliable candidates, mainly for an estimate of the redshift of the arc, in order to confirm or discard their lensing nature.

\section{Conclusions}\label{SECconclusions}

We have developed a new pipeline, based on a Convolutional Neural Network (CNN), to automatically identify strong gravitational lens candidates based mainly on their morphology (Sect.~\ref{SECtraining}). We have applied the method (see Sect.~\ref{SECresults}) to the third data release of the Kilo Degree Survey (KiDS), which is one of the ESO public surveys carried out with the VLT Survey Telescope \cite[see e.g.][for a description]{deJong+15_KiDS_paperI}. Thanks to its high quality images, KiDS is particularly suitable for a search of strong lenses. In the complete survey we expect to find at least 100 LRG lens systems with lens-galaxies at $z<0.4$, possibly increasing to several thousand when expanding the search to fainter and higher-redshift galaxies (see Sect.~\ref{sec:expected}).

To train the CNN to find lenses, we generated a large sample of simulated lensing features on top of observed colour-magnitude selected  galaxies from KiDS (Sect.~\ref{sec:sims}). The trained network has been applied to a sample of 21789 LRGs
in KiDS DR3, retrieving 761 candidates ($3.6\%$ of the initial sample). With a visual inspection performed by seven ``human" classifiers, we down-selected the most promising 56 lens candidates (Fig.~\ref{FIGCandidatecutouts}). In our starting sample there were three known lenses, two of which were classified correctly as lenses by the CNN and in the subsequent visual inspection phase (Fig.~\ref{FIGknownlenses}). The misclassified lens has an Einstein radius (0.83 arcsec), well below the range where the CNN is trained ($1.4-5.0$ arcsec). 

For the candidates with available measures of velocity dispersion or stellar mass estimates, we performed an additional sanity check, suggesting that $\sim 22$ are solid candidates (Sect.~\ref{SECradiuscheck}). Considering the colour-magnitude selection of the lens-galaxy sample, the type of lenses simulated and the completeness, the number is roughly consistent with the expected $\sim50$ lenses forecast for the KiDS-DR3 survey area  (\Sec\ref{sec:expected}). Extending this result to the full KiDS survey, we expect to find $\sim100$ LRG lens candidates as a lower limit, similar to the number of lenses in the SLACS sample.
Because we limited our search to a very restricted portion of the colour-magnitude and Einstein-radius and magnification space, dominated by luminous lenses and highly magnified sources,  the natural next step is to enlarge the colour-magnitude pre-selection  of the simulated and observed lenses, which would allow in principle to find up to $\sim2400$ lenses in the most optimistic scenario (nearly all lenses with signal-to-noise larger than 20 and at least three PSF resolution element for the arc-like images). 

A critical aspect to improve in the CNN approach is to reduce the contamination by the false positives which dominate the number of true positives currently by a factor of $\sim 40$. This could facilitate, or even eliminate, the need for visual inspection.
This factor is consistent with the $\sim 4\%$ mis-classifications
in the training (see Appendix~\ref{SEC:appendix}), which for the 
input sample size can lead to $\sim 900$ false-positive, which 
is close to the actual number of $\sim 700$. Given that the 
lenses-galaxies are outnumbered by normal galaxies typically by a thousand to one, an important goal is to bring the false-positive rate down to less than 0.1\%, without decreasing the true-positive rate substantially. This is a hard task, but human visual inspection suggests that at least 0.25\% can be reached (i.e. 56 out of 22 thousand), possibly when including additional colour (RGB) information. Our next goal is to create a completely automated pipeline for lens classification without the need of visual inspection. However, if visual inspection will be needed to down-select lens candidates, it will be important to test the efficiency of the human classifiers in order to be able to estimate accurately the completeness and purity of the final selected sample.

Moreover, for evaluating purity and completeness in a realistic lens search setting, we plan to build a validation-set which reproduces the characteristics of a real survey (where the number of negatives far outnumber the number of positives).

The CNN tends to mis-classify primarily galaxies resembling lensing features (e.g., ring galaxies, mergers, star-forming rings). Thus, training the network on an ensemble of this kind of false-positives would allow the algorithm to learn the subtle differences between the false positives and the true lenses.

The network performance could also be improved with model averaging, i.e., building a series of networks for the same task, but with different structure and parameters, and by averaging their output. Moreover, training the network on galaxy-subtracted images could facilitate the algorithm to pick up more subtle lensing features, especially in the regime of small Einstein radii and bright galaxies. Another possibility is to produce and train the CNN on multi-band images. In this way, colour information would be used to discriminate between lenses with sources and non-lenses. 

The final sample of KiDS-DR3 lens candidates suggests that our method is promising to down-selected lens candidates from an input sample by two orders of magnitude. Moreover, it is easily applicable to any ongoing and future survey, (e.g., Euclid, LSST) for classyfing the enormous amount of data that will be produced. In the near future we plan  spectroscopic follow-up of our best candidates, to model them, and to better assess their selection biases. In addition, we will apply the method to the full KiDS survey and work on the above-mentioned improvements.

\vfill

\captionsetup[subfigure]{labelformat=empty}
 \begin{figure*}
   \centering
   \subfloat[KSL427 (70)]{\includegraphics[width=38mm]{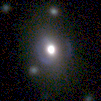}}\hspace{\fill}
   \subfloat[KSL317 (70)]{\includegraphics[width=38mm]{rgb2.png}}\hspace{\fill}
   \subfloat[KSL103 (64)]{\includegraphics[width=38mm]{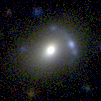}}\hspace{\fill}
   \subfloat[KSL627 (60)]{\includegraphics[width=38mm]{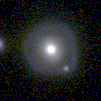}}\hspace{\fill}
   \subfloat[KSL040 (60)]{\includegraphics[width=38mm]{rgb5.png}}\hspace{\fill}
   \subfloat[KSL327 (58)]{\includegraphics[width=38mm]{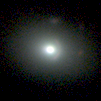}}\hspace{\fill}
   \subfloat[KSL376 (48)]{\includegraphics[width=38mm]{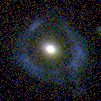}}\hspace{\fill}
   \subfloat[KSL086 (48)]{\includegraphics[width=38mm]{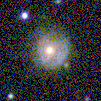}}\hspace{\fill}
   \subfloat[KSL469 (46)]{\includegraphics[width=38mm]{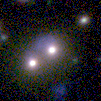}}\hspace{\fill}
   \subfloat[KSL351 (46)]{\includegraphics[width=38mm]{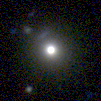}}\hspace{\fill}
   \subfloat[KSL713 (42)]{\includegraphics[width=38mm]{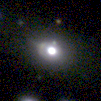}}\hspace{\fill}
   \subfloat[KSL328 (42)]{\includegraphics[width=38mm]{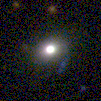}}\hspace{\fill}
   \subfloat[KSL228 (42)]{\includegraphics[width=38mm]{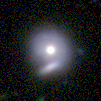}}\hspace{\fill}
   \subfloat[KSL411 (40)]{\includegraphics[width=38mm]{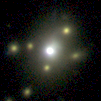}}\hspace{\fill}
   \subfloat[KSL070 (40)]{\includegraphics[width=38mm]{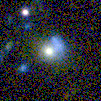}}\hspace{\fill}
   \subfloat[KSL543 (38)]{\includegraphics[width=38mm]{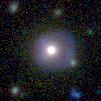}}
\caption{RGB images of the 56 candidates down-selected through a visual inspection of the 761 CNN candidates (see \Sec\ref{SECvisual}). Each source is labelled by an internal ID followed by, in parentheses, the visual classification score (70 points maximum). Each image is 20 by 20 arcsec.}
 \label{FIGCandidatecutouts}
 \end{figure*}

\begin{figure*}
  \centering
  \subfloat[KSL664 (36)]{\includegraphics[width=38mm]{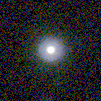}}\hspace{\fill}
  \subfloat[KSL106 (36)]{\includegraphics[width=38mm]{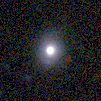}}\hspace{\fill}
  \subfloat[KSL415 (32)]{\includegraphics[width=38mm]{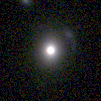}}\hspace{\fill}
  \subfloat[KSL388 (32)]{\includegraphics[width=38mm]{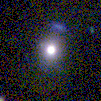}}\hspace{\fill}
  \subfloat[KSL337 (32)]{\includegraphics[width=38mm]{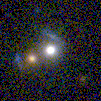}}\hspace{\fill}
  \subfloat[KSL220 (30)]{\includegraphics[width=38mm]{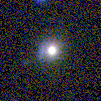}}\hspace{\fill}
  \subfloat[KSL603 (28)]{\includegraphics[width=38mm]{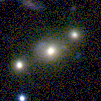}}\hspace{\fill}
  \subfloat[KSL601 (28)]{\includegraphics[width=38mm]{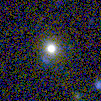}}\hspace{\fill}
  \subfloat[KSL737 (26)]{\includegraphics[width=38mm]{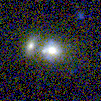}}\hspace{\fill}
  \subfloat[KSL669 (26)]{\includegraphics[width=38mm]{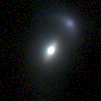}}\hspace{\fill}
  \subfloat[KSL450 (26)]{\includegraphics[width=38mm]{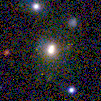}}\hspace{\fill}
  \subfloat[KSL436 (26)]{\includegraphics[width=38mm]{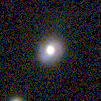}}\hspace{\fill}
  \subfloat[KSL233 (26)]{\includegraphics[width=38mm]{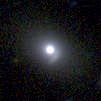}}\hspace{\fill}
  \subfloat[KSL231 (26)]{\includegraphics[width=38mm]{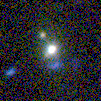}}\hspace{\fill}
  \subfloat[KSL101 (26)]{\includegraphics[width=38mm]{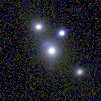}}\hspace{\fill}
  \subfloat[KSL094 (26)]{\includegraphics[width=38mm]{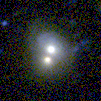}}\hspace{\fill}
  \subfloat[KSL707 (24)]{\includegraphics[width=38mm]{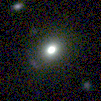}}\hspace{\fill}
  \subfloat[KSL565 (24)]{\includegraphics[width=38mm]{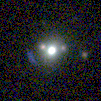}}\hspace{\fill}
  \subfloat[KSL335 (24)]{\includegraphics[width=38mm]{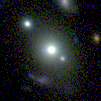}}\hspace{\fill}
  \subfloat[KSL197 (24)]{\includegraphics[width=38mm]{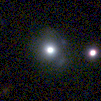}}
\contcaption{}
\end{figure*}

\begin{figure*}
  \centering
  \subfloat[KSL134 (24)]{\includegraphics[width=38mm]{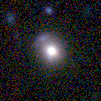}}\hspace{\fill}
  \subfloat[KSL620 (22)]{\includegraphics[width=38mm]{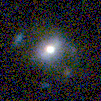}}\hspace{\fill}
  \subfloat[KSL606 (22)]{\includegraphics[width=38mm]{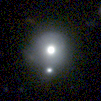}}\hspace{\fill}
  \subfloat[KSL434 (22)]{\includegraphics[width=38mm]{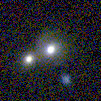}}\hspace{\fill}
  \subfloat[KSL421 (22)]{\includegraphics[width=38mm]{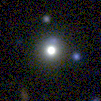}}\hspace{\fill}
  \subfloat[KSL046 (22)]{\includegraphics[width=38mm]{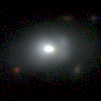}}\hspace{\fill}
  \subfloat[KSL013 (22)]{\includegraphics[width=38mm]{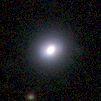}}\hspace{\fill}
  \subfloat[KSL686 (20)]{\includegraphics[width=38mm]{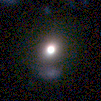}}\hspace{\fill}
  \subfloat[KSL674 (20)]{\includegraphics[width=38mm]{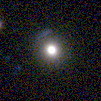}}\hspace{\fill}
  \subfloat[KSL670 (20)]{\includegraphics[width=38mm]{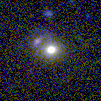}}\hspace{\fill}
  \subfloat[KSL564 (20)]{\includegraphics[width=38mm]{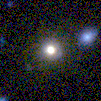}}\hspace{\fill}
  \subfloat[KSL516 (20)]{\includegraphics[width=38mm]{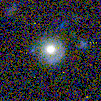}}\hspace{\fill}
  \subfloat[KSL465 (20)]{\includegraphics[width=38mm]{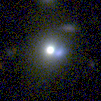}}\hspace{\fill}
  \subfloat[KSL463 (20)]{\includegraphics[width=38mm]{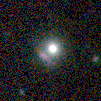}}\hspace{\fill}
  \subfloat[KSL342 (20)]{\includegraphics[width=38mm]{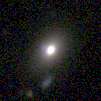}}\hspace{\fill}
  \subfloat[KSL322 (20)]{\includegraphics[width=38mm]{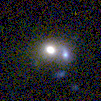}}\hspace{\fill}
  \subfloat[KSL278 (20)]{\includegraphics[width=38mm]{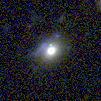}}\hspace{\fill}
  \subfloat[KSL178 (20)]{\includegraphics[width=38mm]{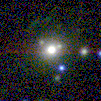}}\hspace{\fill}
  \subfloat[KSL159 (20)]{\includegraphics[width=38mm]{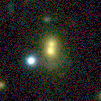}}\hspace{\fill}
  \subfloat[KSL535 (18)]{\includegraphics[width=38mm]{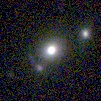}}
  \contcaption{}
\end{figure*}

\section*{acknowledgements}

We thank the anonymous referee for his/her comments that have improved the paper; Sander Dieleman, Davide Punzo, Marco Alexander Wiering, Emmanuel Okafor, Giuseppe Longo, Andrea Colonna and Koen Kuijken for stimulating and useful discussions; Hugo Buddelmeijer, Ewout Helmich and Jelte de Jong for their technical support. LVEK thanks Tom Collett with help in using the code \textsc{LensPop}. CEP, SC, CT, GV, GVK and LVEK are supported through an NWO-VICI grant (project number 639.043.308).
GVK acknowledges financial support from the Netherlands Research School for Astronomy (NOVA) and Target. Target is supported by Samenwerkingsverband Noord Nederland, European fund for regional development, Dutch Ministry of economic affairs, Pieken in de Delta, Provinces of Groningen and Drenthe. PS is supported by the Deutsche Forschungsgemeinschaft in the framework of the TR33 `The Dark Universe'.

\bibliographystyle{mnras}
\bibliography{petrillo}

\begin{thebibliography}{}
\makeatletter
\relax
\def\mn@urlcharsother{\let\do\@makeother \do\$\do\&\do\#\do\^\do\_\do\%\do\~}
\def\mn@doi{\begingroup\mn@urlcharsother \@ifnextchar [ {\mn@doi@}
  {\mn@doi@[]}}
\def\mn@doi@[#1]#2{\def\@tempa{#1}\ifx\@tempa\@empty \href
  {http://dx.doi.org/#2} {doi:#2}\else \href {http://dx.doi.org/#2} {#1}\fi
  \endgroup}
\def\mn@eprint#1#2{\mn@eprint@#1:#2::\@nil}
\def\mn@eprint@arXiv#1{\href {http://arxiv.org/abs/#1} {{\tt arXiv:#1}}}
\def\mn@eprint@dblp#1{\href {http://dblp.uni-trier.de/rec/bibtex/#1.xml}
  {dblp:#1}}
\def\mn@eprint@#1:#2:#3:#4\@nil{\def\@tempa {#1}\def\@tempb {#2}\def\@tempc
  {#3}\ifx \@tempc \@empty \let \@tempc \@tempb \let \@tempb \@tempa \fi \ifx
  \@tempb \@empty \def\@tempb {arXiv}\fi \@ifundefined
  {mn@eprint@\@tempb}{\@tempb:\@tempc}{\expandafter \expandafter \csname
  mn@eprint@\@tempb\endcsname \expandafter{\@tempc}}}

\bibitem[\protect\citeauthoryear{{Alard}}{{Alard}}{2006}]{Alard2006}
{Alard} C.,  2006, preprint, \href
  {http://adsabs.harvard.edu/abs/2006astro.ph..6757A} {} (\mn@eprint {arXiv}
  {astro-ph/0606757})

\bibitem[\protect\citeauthoryear{{Arnouts}, {Cristiani}, {Moscardini},
  {Matarrese}, {Lucchin}, {Fontana}  \& {Giallongo}}{{Arnouts}
  et~al.}{1999}]{Arnouts+99}
{Arnouts} S.,  {Cristiani} S.,  {Moscardini} L.,  {Matarrese} S.,  {Lucchin}
  F.,  {Fontana} A.,   {Giallongo} E.,  1999, \mn@doi [\mnras]
  {10.1046/j.1365-8711.1999.02978.x}, \href
  {http://adsabs.harvard.edu/abs/1999MNRAS.310..540A} {310, 540}

\bibitem[\protect\citeauthoryear{{Auger}, {Treu}, {Bolton}, {Gavazzi},
  {Koopmans}, {Marshall}, {Bundy}  \& {Moustakas}}{{Auger}
  et~al.}{2009}]{Auger+09_SLACSIX}
{Auger} M.~W.,  {Treu} T.,  {Bolton} A.~S.,  {Gavazzi} R.,  {Koopmans}
  L.~V.~E.,  {Marshall} P.~J.,  {Bundy} K.,   {Moustakas} L.~A.,  2009, \mn@doi
  [\apj] {10.1088/0004-637X/705/2/1099}, \href
  {http://adsabs.harvard.edu/abs/2009ApJ...705.1099A} {705, 1099}

\bibitem[\protect\citeauthoryear{{Barnab{\`e}}, {Czoske}, {Koopmans}, {Treu},
  {Bolton}  \& {Gavazzi}}{{Barnab{\`e}} et~al.}{2009}]{barnabe2009}
{Barnab{\`e}} M.,  {Czoske} O.,  {Koopmans} L.~V.~E.,  {Treu} T.,  {Bolton}
  A.~S.,   {Gavazzi} R.,  2009, \mn@doi [\mnras]
  {10.1111/j.1365-2966.2009.14941.x}, \href
  {http://adsabs.harvard.edu/abs/2009MNRAS.399...21B} {399, 21}

\bibitem[\protect\citeauthoryear{{Barnab{\`e}}, {Czoske}, {Koopmans}, {Treu}
  \& {Bolton}}{{Barnab{\`e}} et~al.}{2011}]{Barnab2011}
{Barnab{\`e}} M.,  {Czoske} O.,  {Koopmans} L.~V.~E.,  {Treu} T.,   {Bolton}
  A.~S.,  2011, \mn@doi [\mnras] {10.1111/j.1365-2966.2011.18842.x}, \href
  {http://adsabs.harvard.edu/abs/2011MNRAS.415.2215B} {415, 2215}

\bibitem[\protect\citeauthoryear{{Ben{\'{\i}}tez}}{{Ben{\'{\i}}tez}}{2000}]{Benitez00}
{Ben{\'{\i}}tez} N.,  2000, \mn@doi [\apj] {10.1086/308947}, \href
  {http://adsabs.harvard.edu/abs/2000ApJ...536..571B} {536, 571}

\bibitem[\protect\citeauthoryear{{Bertin} \& {Arnouts}}{{Bertin} \&
  {Arnouts}}{1996}]{Bertin_Arnouts96_SEx}
{Bertin} E.,  {Arnouts} S.,  1996, \mn@doi [\aaps] {10.1051/aas:1996164}, \href
  {http://adsabs.harvard.edu/abs/1996A%26AS..117..393B} {117, 393}

\bibitem[\protect\citeauthoryear{{Binney} \& {Merrifield}}{{Binney} \&
  {Merrifield}}{1998}]{Binney_Merrifield98_book}
{Binney} J.,  {Merrifield} M.,  1998, {Galactic Astronomy}.
Princeton University Press

\bibitem[\protect\citeauthoryear{{Bolton}, {Burles}, {Koopmans}, {Treu},
  {Gavazzi}, {Moustakas}, {Wayth}  \& {Schlegel}}{{Bolton}
  et~al.}{2008}]{bolton2008}
{Bolton} A.~S.,  {Burles} S.,  {Koopmans} L.~V.~E.,  {Treu} T.,  {Gavazzi} R.,
  {Moustakas} L.~A.,  {Wayth} R.,   {Schlegel} D.~J.,  2008, \mn@doi [\apj]
  {10.1086/589327}, \href {http://adsabs.harvard.edu/abs/2008ApJ...682..964B}
  {682, 964}

\bibitem[\protect\citeauthoryear{{Bom}, {Makler}, {Albuquerque}  \&
  {Brandt}}{{Bom} et~al.}{2016}]{Bom2016}
{Bom} C.~R.,  {Makler} M.,  {Albuquerque} M.~P.,   {Brandt} C.~H.,  2016,
  preprint, \href {http://adsabs.harvard.edu/abs/2016arXiv160704644B} {}
  (\mn@eprint {arXiv} {1607.04644})

\bibitem[\protect\citeauthoryear{{Bonvin} et~al.,}{{Bonvin}
  et~al.}{2016}]{bonvin2016}
{Bonvin} V.,  et~al., 2016, preprint, \href
  {http://adsabs.harvard.edu/abs/2016arXiv160701790B} {} (\mn@eprint {arXiv}
  {1607.01790})

\bibitem[\protect\citeauthoryear{Boureau, Bach, LeCun  \& Ponce}{Boureau
  et~al.}{2010}]{boureau2010learning}
Boureau Y.-L.,  Bach F.,  LeCun Y.,   Ponce J.,  2010, in Computer Vision and
  Pattern Recognition (CVPR), 2010 IEEE Conference on. pp 2559--2566

\bibitem[\protect\citeauthoryear{Bournaud \& Combes}{Bournaud \&
  Combes}{2003}]{bournaud2003}
Bournaud F.,  Combes F.,  2003, \aap, 401, 817

\bibitem[\protect\citeauthoryear{{Brault} \& {Gavazzi}}{{Brault} \&
  {Gavazzi}}{2015}]{Brault2015}
{Brault} F.,  {Gavazzi} R.,  2015, \mn@doi [\aap]
  {10.1051/0004-6361/201425275}, \href
  {http://adsabs.harvard.edu/abs/2015A%26A...577A..85B} {577, A85}

\bibitem[\protect\citeauthoryear{Brewer et~al.,}{Brewer
  et~al.}{2012}]{brewer2012swells}
Brewer B.~J.,  et~al., 2012, \mnras, 422, 3574

\bibitem[\protect\citeauthoryear{{Bruzual} \& {Charlot}}{{Bruzual} \&
  {Charlot}}{2003}]{BC03}
{Bruzual} G.,  {Charlot} S.,  2003, \mn@doi [\mnras]
  {10.1046/j.1365-8711.2003.06897.x}, \href
  {http://adsabs.harvard.edu/abs/2003MNRAS.344.1000B} {344, 1000}

\bibitem[\protect\citeauthoryear{{Cao}, {Biesiada}, {Yao}  \& {Zhu}}{{Cao}
  et~al.}{2016}]{cao2016limits}
{Cao} S.,  {Biesiada} M.,  {Yao} M.,   {Zhu} Z.-H.,  2016, \mn@doi [\mnras]
  {10.1093/mnras/stw932}, \href
  {http://adsabs.harvard.edu/abs/2016MNRAS.461.2192C} {461, 2192}

\bibitem[\protect\citeauthoryear{{Capaccioli} \& {Schipani}}{{Capaccioli} \&
  {Schipani}}{2011}]{Capaccioli_Schipani11}
{Capaccioli} M.,  {Schipani} P.,  2011, The Messenger, \href
  {http://adsabs.harvard.edu/abs/2011Msngr.146....2C} {146, 2}

\bibitem[\protect\citeauthoryear{{Cardone} \& {Tortora}}{{Cardone} \&
  {Tortora}}{2010}]{Cardone2010}
{Cardone} V.~F.,  {Tortora} C.,  2010, \mn@doi [\mnras]
  {10.1111/j.1365-2966.2010.17398.x}, \href
  {http://adsabs.harvard.edu/abs/2010MNRAS.409.1570C} {409, 1570}

\bibitem[\protect\citeauthoryear{{Cardone}, {Tortora}, {Molinaro}  \&
  {Salzano}}{{Cardone} et~al.}{2009}]{cardone2009}
{Cardone} V.~F.,  {Tortora} C.,  {Molinaro} R.,   {Salzano} V.,  2009, \mn@doi
  [\aap] {10.1051/0004-6361/200811090}, \href
  {http://adsabs.harvard.edu/abs/2009A%26A...504..769C} {504, 769}

\bibitem[\protect\citeauthoryear{{Carlstrom} et~al.,}{{Carlstrom}
  et~al.}{2011}]{carlstrom2011}
{Carlstrom} J.~E.,  et~al., 2011, \mn@doi [\pasp] {10.1086/659879}, \href
  {http://adsabs.harvard.edu/abs/2011PASP..123..568C} {123, 568}

\bibitem[\protect\citeauthoryear{{Chabrier}}{{Chabrier}}{2001}]{Chabrier01}
{Chabrier} G.,  2001, \mn@doi [\apj] {10.1086/321401}, \href
  {http://adsabs.harvard.edu/abs/2001ApJ...554.1274C} {554, 1274}

\bibitem[\protect\citeauthoryear{Chae}{Chae}{2003}]{chae2003cosmic}
Chae K.-H.,  2003, \mnras, 346, 746

\bibitem[\protect\citeauthoryear{{Chevance}, {Weijmans}, {Damjanov}, {Abraham},
  {Simard}, {van den Bergh}, {Caris}  \& {Glazebrook}}{{Chevance}
  et~al.}{2012}]{Chevance+12}
{Chevance} M.,  {Weijmans} A.-M.,  {Damjanov} I.,  {Abraham} R.~G.,  {Simard}
  L.,  {van den Bergh} S.,  {Caris} E.,   {Glazebrook} K.,  2012, \mn@doi
  [\apjl] {10.1088/2041-8205/754/2/L24}, \href
  {http://adsabs.harvard.edu/abs/2012ApJ...754L..24C} {754, L24}

\bibitem[\protect\citeauthoryear{Colless et~al.,}{Colless
  et~al.}{2001}]{2dfColless2001}
Colless M.,  et~al., 2001, \mnras, 328, 1039

\bibitem[\protect\citeauthoryear{{Collett}}{{Collett}}{2015}]{collet2015}
{Collett} T.~E.,  2015, \mn@doi [\apj] {10.1088/0004-637X/811/1/20}, \href
  {http://adsabs.harvard.edu/abs/2015ApJ...811...20C} {811, 20}

\bibitem[\protect\citeauthoryear{{Dawson} et~al.,}{{Dawson}
  et~al.}{2013}]{BOSS}
{Dawson} K.~S.,  et~al., 2013, \mn@doi [\aj] {10.1088/0004-6256/145/1/10},
  \href {http://adsabs.harvard.edu/abs/2013AJ....145...10D} {145, 10}

\bibitem[\protect\citeauthoryear{{Deane}, {Rawlings}, {Garrett}, {Heywood},
  {Jarvis}, {Kl{\"o}ckner}, {Marshall}  \& {McKean}}{{Deane}
  et~al.}{2013}]{deane2013}
{Deane} R.~P.,  {Rawlings} S.,  {Garrett} M.~A.,  {Heywood} I.,  {Jarvis}
  M.~J.,  {Kl{\"o}ckner} H.-R.,  {Marshall} P.~J.,   {McKean} J.~P.,  2013,
  \mn@doi [\mnras] {10.1093/mnras/stt1241}, \href
  {http://adsabs.harvard.edu/abs/2013MNRAS.434.3322D} {434, 3322}

\bibitem[\protect\citeauthoryear{Dieleman, Willett  \& Dambre}{Dieleman
  et~al.}{2015}]{dieleman2015rotation}
Dieleman S.,  Willett K.~W.,   Dambre J.,  2015, \mnras, 450, 1441

\bibitem[\protect\citeauthoryear{{Eisenstein} et~al.,}{{Eisenstein}
  et~al.}{2001}]{Eisenstein2001}
{Eisenstein} D.~J.,  et~al., 2001, \mn@doi [\aj] {10.1086/323717}, \href
  {http://adsabs.harvard.edu/abs/2001AJ....122.2267E} {122, 2267}

\bibitem[\protect\citeauthoryear{{Eisenstein} et~al.,}{{Eisenstein}
  et~al.}{2011}]{SDSS}
{Eisenstein} D.~J.,  et~al., 2011, \mn@doi [\aj] {10.1088/0004-6256/142/3/72},
  \href {http://adsabs.harvard.edu/abs/2011AJ....142...72E} {142, 72}

\bibitem[\protect\citeauthoryear{{Estrada} et~al.,}{{Estrada}
  et~al.}{2007}]{Estrada2007}
{Estrada} J.,  et~al., 2007, \mn@doi [\apj] {10.1086/512599}, \href
  {http://adsabs.harvard.edu/abs/2007ApJ...660.1176E} {660, 1176}

\bibitem[\protect\citeauthoryear{{Ferreras}, {Saha}, {Leier}, {Courbin}  \&
  {Falco}}{{Ferreras} et~al.}{2010}]{ferreras2010}
{Ferreras} I.,  {Saha} P.,  {Leier} D.,  {Courbin} F.,   {Falco} E.~E.,  2010,
  \mn@doi [\mnras] {10.1111/j.1745-3933.2010.00941.x}, \href
  {http://adsabs.harvard.edu/abs/2010MNRAS.409L..30F} {409, L30}

\bibitem[\protect\citeauthoryear{{Fo{\"e}x}, {Motta}, {Limousin}, {Verdugo},
  {More}, {Cabanac}, {Gavazzi}  \& {Mu{\~n}oz}}{{Fo{\"e}x}
  et~al.}{2013}]{Foex+13_SARCS}
{Fo{\"e}x} G.,  {Motta} V.,  {Limousin} M.,  {Verdugo} T.,  {More} A.,
  {Cabanac} R.,  {Gavazzi} R.,   {Mu{\~n}oz} R.~P.,  2013, \mn@doi [\aap]
  {10.1051/0004-6361/201321112}, \href
  {http://adsabs.harvard.edu/abs/2013A%26A...559A.105F} {559, A105}

\bibitem[\protect\citeauthoryear{Fukugita, Futamase, Kasai  \& Turner}{Fukugita
  et~al.}{1992}]{fukugita1992statistical}
Fukugita M.,  Futamase T.,  Kasai M.,   Turner E.,  1992, \apj, 393, 3

\bibitem[\protect\citeauthoryear{Fukushima}{Fukushima}{1980}]{fukushima1980neocognitron}
Fukushima K.,  1980, Biological cybernetics, 36, 193

\bibitem[\protect\citeauthoryear{{Gavazzi}, {Treu}, {Rhodes}, {Koopmans},
  {Bolton}, {Burles}, {Massey}  \& {Moustakas}}{{Gavazzi}
  et~al.}{2007}]{gavazzi2007}
{Gavazzi} R.,  {Treu} T.,  {Rhodes} J.~D.,  {Koopmans} L.~V.~E.,  {Bolton}
  A.~S.,  {Burles} S.,  {Massey} R.~J.,   {Moustakas} L.~A.,  2007, \mn@doi
  [\apj] {10.1086/519237}, \href
  {http://adsabs.harvard.edu/abs/2007ApJ...667..176G} {667, 176}

\bibitem[\protect\citeauthoryear{{Gavazzi}, {Marshall}, {Treu}  \&
  {Sonnenfeld}}{{Gavazzi} et~al.}{2014}]{Gavazzi2014}
{Gavazzi} R.,  {Marshall} P.~J.,  {Treu} T.,   {Sonnenfeld} A.,  2014, \mn@doi
  [\apj] {10.1088/0004-637X/785/2/144}, \href
  {http://adsabs.harvard.edu/abs/2014ApJ...785..144G} {785, 144}

\bibitem[\protect\citeauthoryear{{Grillo}, {Eichner}, {Seitz}, {Bender},
  {Lombardi}, {Gobat}  \& {Bauer}}{{Grillo} et~al.}{2010}]{grillo2010}
{Grillo} C.,  {Eichner} T.,  {Seitz} S.,  {Bender} R.,  {Lombardi} M.,  {Gobat}
  R.,   {Bauer} A.,  2010, \mn@doi [\apj] {10.1088/0004-637X/710/1/372}, \href
  {http://adsabs.harvard.edu/abs/2010ApJ...710..372G} {710, 372}

\bibitem[\protect\citeauthoryear{{Grogin} et~al.,}{{Grogin}
  et~al.}{2011}]{CANDELS}
{Grogin} N.~A.,  et~al., 2011, \mn@doi [\apjs] {10.1088/0067-0049/197/2/35},
  \href {http://adsabs.harvard.edu/abs/2011ApJS..197...35G} {197, 35}

\bibitem[\protect\citeauthoryear{Guo, Liu, Oerlemans, Lao, Wu  \& Lew}{Guo
  et~al.}{2016}]{guo2016deep}
Guo Y.,  Liu Y.,  Oerlemans A.,  Lao S.,  Wu S.,   Lew M.~S.,  2016,
  Neurocomputing, 187, 27

\bibitem[\protect\citeauthoryear{{H{\'a}la}}{{H{\'a}la}}{2014}]{Hala2014}
{H{\'a}la} P.,  2014, preprint, \href
  {http://adsabs.harvard.edu/abs/2014arXiv1412.8341H} {} (\mn@eprint {arXiv}
  {1412.8341})

\bibitem[\protect\citeauthoryear{He, Zhang, Ren  \& Sun}{He
  et~al.}{2015a}]{he2015delving}
He K.,  Zhang X.,  Ren S.,   Sun J.,  2015a, in Proceedings of the IEEE
  International Conference on Computer Vision. pp 1026--1034

\bibitem[\protect\citeauthoryear{He, Zhang, Ren  \& Sun}{He
  et~al.}{2015b}]{he2015deep}
He K.,  Zhang X.,  Ren S.,   Sun J.,  2015b, CoRR, abs/1512.03385

\bibitem[\protect\citeauthoryear{{Hinton}, {Srivastava}, {Krizhevsky},
  {Sutskever}  \& {Salakhutdinov}}{{Hinton} et~al.}{2012}]{hinton2012improving}
{Hinton} G.~E.,  {Srivastava} N.,  {Krizhevsky} A.,  {Sutskever} I.,
  {Salakhutdinov} R.~R.,  2012, preprint, \href
  {http://adsabs.harvard.edu/abs/2012arXiv1207.0580H} {} (\mn@eprint {arXiv}
  {1207.0580})

\bibitem[\protect\citeauthoryear{{Hoag}}{{Hoag}}{1950}]{Hoag1950}
{Hoag} A.~A.,  1950, \mn@doi [\aj] {10.1086/106427}, \href
  {http://adsabs.harvard.edu/abs/1950AJ.....55Q.170H} {55, 170}

\bibitem[\protect\citeauthoryear{{Horesh}, {Ofek}, {Maoz}, {Bartelmann},
  {Meneghetti}  \& {Rix}}{{Horesh} et~al.}{2005}]{Horesh2005}
{Horesh} A.,  {Ofek} E.~O.,  {Maoz} D.,  {Bartelmann} M.,  {Meneghetti} M.,
  {Rix} H.-W.,  2005, \mn@doi [\apj] {10.1086/466519}, \href
  {http://adsabs.harvard.edu/abs/2005ApJ...633..768H} {633, 768}

\bibitem[\protect\citeauthoryear{{Hoyle}}{{Hoyle}}{2016}]{Hoyle2016}
{Hoyle} B.,  2016, \mn@doi [Astronomy and Computing]
  {10.1016/j.ascom.2016.03.006}, \href
  {http://adsabs.harvard.edu/abs/2016A%26C....16...34H} {16, 34}

\bibitem[\protect\citeauthoryear{{Huertas-Company} et~al.,}{{Huertas-Company}
  et~al.}{2015}]{Huertas2015}
{Huertas-Company} M.,  et~al., 2015, \mn@doi [\apjs]
  {10.1088/0067-0049/221/1/8}, \href
  {http://adsabs.harvard.edu/abs/2015ApJS..221....8H} {221, 8}

\bibitem[\protect\citeauthoryear{{Ilbert} et~al.,}{{Ilbert}
  et~al.}{2006}]{Ilbert+06}
{Ilbert} O.,  et~al., 2006, \mn@doi [\aap] {10.1051/0004-6361:20065138}, \href
  {http://adsabs.harvard.edu/abs/2006A%26A...457..841I} {457, 841}

\bibitem[\protect\citeauthoryear{{Impellizzeri}, {McKean}, {Castangia}, {Roy},
  {Henkel}, {Brunthaler}  \& {Wucknitz}}{{Impellizzeri}
  et~al.}{2008}]{impellizzeri2008}
{Impellizzeri} C.~M.~V.,  {McKean} J.~P.,  {Castangia} P.,  {Roy} A.~L.,
  {Henkel} C.,  {Brunthaler} A.,   {Wucknitz} O.,  2008, \mn@doi [\nat]
  {10.1038/nature07544}, \href
  {http://adsabs.harvard.edu/abs/2008Natur.456..927I} {456, 927}

\bibitem[\protect\citeauthoryear{{Iodice}, {Arnaboldi}, {Bournaud}, {Combes},
  {Sparke}, {van Driel}  \& {Capaccioli}}{{Iodice} et~al.}{2003}]{Iodice2003}
{Iodice} E.,  {Arnaboldi} M.,  {Bournaud} F.,  {Combes} F.,  {Sparke} L.~S.,
  {van Driel} W.,   {Capaccioli} M.,  2003, \mn@doi [\apj] {10.1086/346107},
  \href {http://adsabs.harvard.edu/abs/2003ApJ...585..730I} {585, 730}

\bibitem[\protect\citeauthoryear{{Ioffe} \& {Szegedy}}{{Ioffe} \&
  {Szegedy}}{2015}]{ioffe2015batch}
{Ioffe} S.,  {Szegedy} C.,  2015, preprint, \href
  {http://adsabs.harvard.edu/abs/2015arXiv150203167I} {} (\mn@eprint {arXiv}
  {1502.03167})

\bibitem[\protect\citeauthoryear{{Jiang} \& {Kochanek}}{{Jiang} \&
  {Kochanek}}{2007}]{jiang2007}
{Jiang} G.,  {Kochanek} C.~S.,  2007, \mn@doi [\apj] {10.1086/522580}, \href
  {http://adsabs.harvard.edu/abs/2007ApJ...671.1568J} {671, 1568}

\bibitem[\protect\citeauthoryear{{Joseph} et~al.,}{{Joseph}
  et~al.}{2014}]{Joseph2014}
{Joseph} R.,  et~al., 2014, \mn@doi [\aap] {10.1051/0004-6361/201423365}, \href
  {http://adsabs.harvard.edu/abs/2014A%26A...566A..63J} {566, A63}

\bibitem[\protect\citeauthoryear{{Kim} \& {Brunner}}{{Kim} \&
  {Brunner}}{2016}]{Kim2016}
{Kim} E.~J.,  {Brunner} R.~J.,  2016, preprint, \href
  {http://adsabs.harvard.edu/abs/2016arXiv160804369K} {} (\mn@eprint {arXiv}
  {1608.04369})

\bibitem[\protect\citeauthoryear{{Kingma} \& {Ba}}{{Kingma} \&
  {Ba}}{2014}]{kingma2014adam}
{Kingma} D.,  {Ba} J.,  2014, preprint, \href
  {http://adsabs.harvard.edu/abs/2014arXiv1412.6980K} {} (\mn@eprint {arXiv}
  {1412.6980})

\bibitem[\protect\citeauthoryear{Kochanek}{Kochanek}{1996}]{kochanek1996flat}
Kochanek C.,  1996, \apj, 473, 595

\bibitem[\protect\citeauthoryear{{Koopmans} \& {Treu}}{{Koopmans} \&
  {Treu}}{2003}]{koopmans2003}
{Koopmans} L.~V.~E.,  {Treu} T.,  2003, \mn@doi [\apj] {10.1086/345423}, \href
  {http://adsabs.harvard.edu/abs/2003ApJ...583..606K} {583, 606}

\bibitem[\protect\citeauthoryear{{Koopmans}, {Treu}, {Bolton}, {Burles}  \&
  {Moustakas}}{{Koopmans} et~al.}{2006}]{koopmans2006}
{Koopmans} L.~V.~E.,  {Treu} T.,  {Bolton} A.~S.,  {Burles} S.,   {Moustakas}
  L.~A.,  2006, \mn@doi [\apj] {10.1086/505696}, \href
  {http://adsabs.harvard.edu/abs/2006ApJ...649..599K} {649, 599}

\bibitem[\protect\citeauthoryear{{Koopmans} et~al.,}{{Koopmans}
  et~al.}{2009}]{koopmans2009}
{Koopmans} L.~V.~E.,  et~al., 2009, \mn@doi [\apjl]
  {10.1088/0004-637X/703/1/L51}, \href
  {http://adsabs.harvard.edu/abs/2009ApJ...703L..51K} {703, L51}

\bibitem[\protect\citeauthoryear{{Kormann}, {Schneider}  \&
  {Bartelmann}}{{Kormann} et~al.}{1994}]{KSB_SIE94}
{Kormann} R.,  {Schneider} P.,   {Bartelmann} M.,  1994, \aap, \href
  {http://adsabs.harvard.edu/abs/1994A%26A...284..285K} {284, 285}

\bibitem[\protect\citeauthoryear{{Kubo} \& {Dell'Antonio}}{{Kubo} \&
  {Dell'Antonio}}{2008}]{Kubo2008}
{Kubo} J.~M.,  {Dell'Antonio} I.~P.,  2008, \mn@doi [\mnras]
  {10.1111/j.1365-2966.2008.12880.x}, \href
  {http://adsabs.harvard.edu/abs/2008MNRAS.385..918K} {385, 918}

\bibitem[\protect\citeauthoryear{{Kuijken}}{{Kuijken}}{2011}]{Kuijken11}
{Kuijken} K.,  2011, The Messenger, \href
  {http://adsabs.harvard.edu/abs/2011Msngr.146....8K} {146, 8}

\bibitem[\protect\citeauthoryear{{LSST Science Collaboration} et~al.,}{{LSST
  Science Collaboration} et~al.}{2009}]{abell2009}
{LSST Science Collaboration} et~al., 2009, preprint, \href
  {http://adsabs.harvard.edu/abs/2009arXiv0912.0201L} {} (\mn@eprint {arXiv}
  {0912.0201})

\bibitem[\protect\citeauthoryear{Laureijs et~al.,}{Laureijs
  et~al.}{2011}]{Laureijs:2011wi}
Laureijs R.,  et~al., 2011, arXiv.org

\bibitem[\protect\citeauthoryear{LeCun, Bottou, Bengio  \& Haffner}{LeCun
  et~al.}{1998}]{lecun1998gradient}
LeCun Y.,  Bottou L.,  Bengio Y.,   Haffner P.,  1998, Proceedings of the IEEE,
  86, 2278

\bibitem[\protect\citeauthoryear{LeCun, Bengio  \& Hinton}{LeCun
  et~al.}{2015}]{lecun2015deep}
LeCun Y.,  Bengio Y.,   Hinton G.,  2015, Nature, 521, 436

\bibitem[\protect\citeauthoryear{Leier, Ferreras, Saha, Charlot, Bruzual  \&
  La~Barbera}{Leier et~al.}{2016}]{leier2016strong}
Leier D.,  Ferreras I.,  Saha P.,  Charlot S.,  Bruzual G.,   La~Barbera F.,
  2016, \mnras, 459, 3677

\bibitem[\protect\citeauthoryear{{Lenzen}, {Schindler}  \& {Scherzer}}{{Lenzen}
  et~al.}{2004}]{Lenzen2004}
{Lenzen} F.,  {Schindler} S.,   {Scherzer} O.,  2004, \mn@doi [\aap]
  {10.1051/0004-6361:20034619}, \href
  {http://adsabs.harvard.edu/abs/2004A%26A...416..391L} {416, 391}

\bibitem[\protect\citeauthoryear{{Li}, {Frenk}, {Cole}, {Gao}, {Bose}  \&
  {Hellwing}}{{Li} et~al.}{2016}]{Li2016}
{Li} R.,  {Frenk} C.~S.,  {Cole} S.,  {Gao} L.,  {Bose} S.,   {Hellwing} W.~A.,
   2016, \mn@doi [\mnras] {10.1093/mnras/stw939}, \href
  {http://adsabs.harvard.edu/abs/2016MNRAS.460..363L} {460, 363}

\bibitem[\protect\citeauthoryear{{Limousin} et~al.,}{{Limousin}
  et~al.}{2010}]{Limousin2010}
{Limousin} M.,  et~al., 2010, \mn@doi [\aap] {10.1051/0004-6361/200912747},
  \href {http://adsabs.harvard.edu/abs/2010A%26A...524A..95L} {524, A95}

\bibitem[\protect\citeauthoryear{{Liske} et~al.,}{{Liske}
  et~al.}{2015}]{GAMALiske2015}
{Liske} J.,  et~al., 2015, \mn@doi [\mnras] {10.1093/mnras/stv1436}, \href
  {http://adsabs.harvard.edu/abs/2015MNRAS.452.2087L} {452, 2087}

\bibitem[\protect\citeauthoryear{{Madore}, {Nelson}  \& {Petrillo}}{{Madore}
  et~al.}{2009}]{Madore2009}
{Madore} B.~F.,  {Nelson} E.,   {Petrillo} K.,  2009, \mn@doi [\apjs]
  {10.1088/0067-0049/181/2/572}, \href
  {http://adsabs.harvard.edu/abs/2009ApJS..181..572M} {181, 572}

\bibitem[\protect\citeauthoryear{{Marshall} et~al.,}{{Marshall}
  et~al.}{2016}]{Marshall2016}
{Marshall} P.~J.,  et~al., 2016, \mn@doi [\mnras] {10.1093/mnras/stv2009},
  \href {http://adsabs.harvard.edu/abs/2016MNRAS.455.1171M} {455, 1171}

\bibitem[\protect\citeauthoryear{{Mason} et~al.,}{{Mason}
  et~al.}{2016}]{Mason2016}
{Mason} C.~A.,  et~al., 2016, preprint, \href
  {http://adsabs.harvard.edu/abs/2016arXiv161003075M} {} (\mn@eprint {arXiv}
  {1610.03075})

\bibitem[\protect\citeauthoryear{{Maturi}, {Mizera}  \& {Seidel}}{{Maturi}
  et~al.}{2014}]{Maturi2014}
{Maturi} M.,  {Mizera} S.,   {Seidel} G.,  2014, \mn@doi [\aap]
  {10.1051/0004-6361/201321634}, \href
  {http://adsabs.harvard.edu/abs/2014A%26A...567A.111M} {567, A111}

\bibitem[\protect\citeauthoryear{{McKean} et~al.,}{{McKean}
  et~al.}{2015}]{McKean2015}
{McKean} J.,  et~al., 2015, Proceedings of Science, \href
  {http://adsabs.harvard.edu/abs/2015aska.confE..84M} {p.~84}

\bibitem[\protect\citeauthoryear{Miyazaki et~al.,}{Miyazaki
  et~al.}{2012}]{HSCmiyazaki2012hyper}
Miyazaki S.,  et~al., 2012, in SPIE Astronomical Telescopes+ Instrumentation.
  pp 84460Z--84460Z

\bibitem[\protect\citeauthoryear{M{\"o}ller, Kitzbichler  \&
  Natarajan}{M{\"o}ller et~al.}{2007}]{moller2007strong}
M{\"o}ller O.,  Kitzbichler M.,   Natarajan P.,  2007, \mnras, 379, 1195

\bibitem[\protect\citeauthoryear{{More}, {McKean}, {Muxlow}, {Porcas},
  {Fassnacht}  \& {Koopmans}}{{More} et~al.}{2008}]{Moore2008}
{More} A.,  {McKean} J.~P.,  {Muxlow} T.~W.~B.,  {Porcas} R.~W.,  {Fassnacht}
  C.~D.,   {Koopmans} L.~V.~E.,  2008, \mn@doi [\mnras]
  {10.1111/j.1365-2966.2007.12831.x}, \href
  {http://adsabs.harvard.edu/abs/2008MNRAS.384.1701M} {384, 1701}

\bibitem[\protect\citeauthoryear{{More}, {Jahnke}, {More}, {Gallazzi}, {Bell},
  {Barden}  \& {H{\"a}u{\ss}ler}}{{More} et~al.}{2011}]{more2011}
{More} A.,  {Jahnke} K.,  {More} S.,  {Gallazzi} A.,  {Bell} E.~F.,  {Barden}
  M.,   {H{\"a}u{\ss}ler} B.,  2011, \mn@doi [\apj]
  {10.1088/0004-637X/734/1/69}, \href
  {http://adsabs.harvard.edu/abs/2011ApJ...734...69M} {734, 69}

\bibitem[\protect\citeauthoryear{{More}, {Cabanac}, {More}, {Alard},
  {Limousin}, {Kneib}, {Gavazzi}  \& {Motta}}{{More} et~al.}{2012}]{More2012}
{More} A.,  {Cabanac} R.,  {More} S.,  {Alard} C.,  {Limousin} M.,  {Kneib}
  J.-P.,  {Gavazzi} R.,   {Motta} V.,  2012, \mn@doi [\apj]
  {10.1088/0004-637X/749/1/38}, \href
  {http://adsabs.harvard.edu/abs/2012ApJ...749...38M} {749, 38}

\bibitem[\protect\citeauthoryear{{More} et~al.,}{{More}
  et~al.}{2016}]{more2016}
{More} A.,  et~al., 2016, \mn@doi [\mnras] {10.1093/mnras/stv1965}, \href
  {http://adsabs.harvard.edu/abs/2016MNRAS.455.1191M} {455, 1191}

\bibitem[\protect\citeauthoryear{Nair \& Hinton}{Nair \&
  Hinton}{2010}]{nair2010rectified}
Nair V.,  Hinton G.~E.,  2010, in Proceedings of the 27th International
  Conference on Machine Learning (ICML-10). pp 807--814

\bibitem[\protect\citeauthoryear{{Negrello} et~al.,}{{Negrello}
  et~al.}{2010}]{negrello2010}
{Negrello} M.,  et~al., 2010, \mn@doi [Science] {10.1126/science.1193420},
  \href {http://adsabs.harvard.edu/abs/2010Sci...330..800N} {330, 800}

\bibitem[\protect\citeauthoryear{Ng}{Ng}{2004}]{ng2004feature}
Ng A.~Y.,  2004, in Proceedings of the twenty-first international conference on
  Machine learning. p.~78

\bibitem[\protect\citeauthoryear{Oguri}{Oguri}{2006}]{oguri2006image}
Oguri M.,  2006, \mnras, 367, 1241

\bibitem[\protect\citeauthoryear{Oguri \& Marshall}{Oguri \&
  Marshall}{2010}]{oguri2010gravitationally}
Oguri M.,  Marshall P.~J.,  2010, \mnras, 405, 2579

\bibitem[\protect\citeauthoryear{{Pawase}, {Faure}, {Courbin}, {Kokotanekova}
  \& {Meylan}}{{Pawase} et~al.}{2012}]{pawase2012}
{Pawase} R.~S.,  {Faure} C.,  {Courbin} F.,  {Kokotanekova} R.,   {Meylan} G.,
  2012, preprint, \href {http://adsabs.harvard.edu/abs/2012arXiv1206.3412P} {}
  (\mn@eprint {arXiv} {1206.3412})

\bibitem[\protect\citeauthoryear{Posacki, Cappellari, Treu, Pellegrini  \&
  Ciotti}{Posacki et~al.}{2015}]{posacki2015stellar}
Posacki S.,  Cappellari M.,  Treu T.,  Pellegrini S.,   Ciotti L.,  2015,
  \mnras, 446, 493

\bibitem[\protect\citeauthoryear{{Richard}, {Kneib}, {Ebeling}, {Stark},
  {Egami}  \& {Fiedler}}{{Richard} et~al.}{2011}]{richard2011}
{Richard} J.,  {Kneib} J.-P.,  {Ebeling} H.,  {Stark} D.~P.,  {Egami} E.,
  {Fiedler} A.~K.,  2011, \mn@doi [\mnras] {10.1111/j.1745-3933.2011.01050.x},
  \href {http://adsabs.harvard.edu/abs/2011MNRAS.414L..31R} {414, L31}

\bibitem[\protect\citeauthoryear{{Ruff}, {Gavazzi}, {Marshall}, {Treu}, {Auger}
   \& {Brault}}{{Ruff} et~al.}{2011}]{ruff2011}
{Ruff} A.~J.,  {Gavazzi} R.,  {Marshall} P.~J.,  {Treu} T.,  {Auger} M.~W.,
  {Brault} F.,  2011, \mn@doi [\apj] {10.1088/0004-637X/727/2/96}, \href
  {http://adsabs.harvard.edu/abs/2011ApJ...727...96R} {727, 96}

\bibitem[\protect\citeauthoryear{{Rumelhart}, {Hinton}  \&
  {Williams}}{{Rumelhart} et~al.}{1986}]{backpropagation}
{Rumelhart} D.~E.,  {Hinton} G.~E.,   {Williams} R.~J.,  1986, \mn@doi [\nat]
  {10.1038/323533a0}, \href {http://adsabs.harvard.edu/abs/1986Natur.323..533R}
  {323, 533}

\bibitem[\protect\citeauthoryear{Russakovsky et~al.,}{Russakovsky
  et~al.}{2015}]{ILSVRC15}
Russakovsky O.,  et~al., 2015, \mn@doi [International Journal of Computer
  Vision (IJCV)] {10.1007/s11263-015-0816-y}, 115, 211

\bibitem[\protect\citeauthoryear{{Schlafly} \& {Finkbeiner}}{{Schlafly} \&
  {Finkbeiner}}{2011}]{Schlafly_Finkbeiner11}
{Schlafly} E.~F.,  {Finkbeiner} D.~P.,  2011, \mn@doi [\apj]
  {10.1088/0004-637X/737/2/103}, \href
  {http://adsabs.harvard.edu/abs/2011ApJ...737..103S} {737, 103}

\bibitem[\protect\citeauthoryear{{Schlegel}, {Finkbeiner}  \&
  {Davis}}{{Schlegel} et~al.}{1998}]{SFD98_dust}
{Schlegel} D.~J.,  {Finkbeiner} D.~P.,   {Davis} M.,  1998, \mn@doi [\apj]
  {10.1086/305772}, \href {http://adsabs.harvard.edu/abs/1998ApJ...500..525S}
  {500, 525}

\bibitem[\protect\citeauthoryear{Schmidhuber}{Schmidhuber}{2015}]{schmidhuber2015deep}
Schmidhuber J.,  2015, Neural Networks, 61, 85

\bibitem[\protect\citeauthoryear{Schneider, Ehlers  \& Falco}{Schneider
  et~al.}{1992}]{schneider1992gravitational}
Schneider P.,  Ehlers J.,   Falco E.,  1992, Gravitational Lenses.
Springer-Verlag Berlin Heidelberg New York. Also Astronomy and Astrophysics
  Library

\bibitem[\protect\citeauthoryear{{Seidel} \& {Bartelmann}}{{Seidel} \&
  {Bartelmann}}{2007}]{Seidel2007}
{Seidel} G.,  {Bartelmann} M.,  2007, \mn@doi [\aap]
  {10.1051/0004-6361:20066097}, \href
  {http://adsabs.harvard.edu/abs/2007A%26A...472..341S} {472, 341}

\bibitem[\protect\citeauthoryear{{Sersic}}{{Sersic}}{1968}]{Sersic68}
{Sersic} J.~L.,  1968, {Atlas de galaxias australes}

\bibitem[\protect\citeauthoryear{Simard, Steinkraus  \& Platt}{Simard
  et~al.}{2003}]{Simard2003}
Simard P.~Y.,  Steinkraus D.,   Platt J.~C.,  2003, in Proceedings of the
  Seventh International Conference on Document Analysis and Recognition -
  Volume 2. ICDAR '03.
IEEE Computer Society, Washington, DC, USA, pp 958--, \url
  {http://dl.acm.org/citation.cfm?id=938980.939477}

\bibitem[\protect\citeauthoryear{{Sonnenfeld}, {Treu}, {Gavazzi}, {Suyu},
  {Marshall}, {Auger}  \& {Nipoti}}{{Sonnenfeld}
  et~al.}{2013}]{Sonnenfeld+13_SL2SIV}
{Sonnenfeld} A.,  {Treu} T.,  {Gavazzi} R.,  {Suyu} S.~H.,  {Marshall} P.~J.,
  {Auger} M.~W.,   {Nipoti} C.,  2013, \mn@doi [\apj]
  {10.1088/0004-637X/777/2/98}, \href
  {http://adsabs.harvard.edu/abs/2013ApJ...777...98S} {777, 98}

\bibitem[\protect\citeauthoryear{Sonnenfeld, Treu, Marshall, Suyu, Gavazzi,
  Auger  \& Nipoti}{Sonnenfeld et~al.}{2015}]{sonnenfeld2015sl2s}
Sonnenfeld A.,  Treu T.,  Marshall P.~J.,  Suyu S.~H.,  Gavazzi R.,  Auger
  M.~W.,   Nipoti C.,  2015, \apj, 800, 94

\bibitem[\protect\citeauthoryear{Spiniello, Koopmans, Trager, Czoske  \&
  Treu}{Spiniello et~al.}{2011}]{Spiniello:2011p8239}
Spiniello C.,  Koopmans L.,  Trager S.~C.,  Czoske O.,   Treu T.,  2011, 417,
  3000

\bibitem[\protect\citeauthoryear{{Suyu}, {Marshall}, {Auger}, {Hilbert},
  {Blandford}, {Koopmans}, {Fassnacht}  \& {Treu}}{{Suyu}
  et~al.}{2010}]{suyu2010}
{Suyu} S.~H.,  {Marshall} P.~J.,  {Auger} M.~W.,  {Hilbert} S.,  {Blandford}
  R.~D.,  {Koopmans} L.~V.~E.,  {Fassnacht} C.~D.,   {Treu} T.,  2010, \mn@doi
  [\apj] {10.1088/0004-637X/711/1/201}, \href
  {http://adsabs.harvard.edu/abs/2010ApJ...711..201S} {711, 201}

\bibitem[\protect\citeauthoryear{{Swinbank} et~al.,}{{Swinbank}
  et~al.}{2009}]{swinbank2009}
{Swinbank} A.~M.,  et~al., 2009, \mn@doi [\mnras]
  {10.1111/j.1365-2966.2009.15617.x}, \href
  {http://adsabs.harvard.edu/abs/2009MNRAS.400.1121S} {400, 1121}

\bibitem[\protect\citeauthoryear{{The Dark Energy Survey Collaboration}}{{The
  Dark Energy Survey Collaboration}}{2005}]{DES}
{The Dark Energy Survey Collaboration} 2005, preprint, \href
  {http://adsabs.harvard.edu/abs/2005astro.ph.10346T} {} (\mn@eprint {arXiv}
  {astro-ph/0510346})

\bibitem[\protect\citeauthoryear{{Theano Development Team}}{{Theano Development
  Team}}{2016}]{theano}
{Theano Development Team} 2016, arXiv e-prints, abs/1605.02688

\bibitem[\protect\citeauthoryear{{Theys} \& {Spiegel}}{{Theys} \&
  {Spiegel}}{1976}]{Theys1976}
{Theys} J.~C.,  {Spiegel} E.~A.,  1976, \mn@doi [\apj] {10.1086/154646}, \href
  {http://adsabs.harvard.edu/abs/1976ApJ...208..650T} {208, 650}

\bibitem[\protect\citeauthoryear{{Tortora}, {Napolitano}, {Romanowsky},
  {Capaccioli}  \& {Covone}}{{Tortora} et~al.}{2009}]{Tortora+09}
{Tortora} C.,  {Napolitano} N.~R.,  {Romanowsky} A.~J.,  {Capaccioli} M.,
  {Covone} G.,  2009, \mn@doi [\mnras] {10.1111/j.1365-2966.2009.14789.x},
  \href {http://adsabs.harvard.edu/abs/2009MNRAS.396.1132T} {396, 1132}

\bibitem[\protect\citeauthoryear{Tortora, Napolitano, Romanowsky  \&
  Jetzer}{Tortora et~al.}{2010}]{tortora2010central}
Tortora C.,  Napolitano N.,  Romanowsky A.,   Jetzer P.,  2010, \apj, 721, L1

\bibitem[\protect\citeauthoryear{{Treu} \& {Koopmans}}{{Treu} \&
  {Koopmans}}{2002a}]{Treu2002MNRAS}
{Treu} T.,  {Koopmans} L.~V.~E.,  2002a, \mn@doi [\mnras]
  {10.1046/j.1365-8711.2002.06107.x}, \href
  {http://adsabs.harvard.edu/abs/2002MNRAS.337L...6T} {337, L6}

\bibitem[\protect\citeauthoryear{{Treu} \& {Koopmans}}{{Treu} \&
  {Koopmans}}{2002b}]{Treu2002}
{Treu} T.,  {Koopmans} L.~V.~E.,  2002b, \mn@doi [\apj] {10.1086/341216}, \href
  {http://adsabs.harvard.edu/abs/2002ApJ...575...87T} {575, 87}

\bibitem[\protect\citeauthoryear{{Treu}, {Auger}, {Koopmans}, {Gavazzi},
  {Marshall}  \& {Bolton}}{{Treu} et~al.}{2010}]{treu2010}
{Treu} T.,  {Auger} M.~W.,  {Koopmans} L.~V.~E.,  {Gavazzi} R.,  {Marshall}
  P.~J.,   {Bolton} A.~S.,  2010, \mn@doi [\apj]
  {10.1088/0004-637X/709/2/1195}, \href
  {http://adsabs.harvard.edu/abs/2010ApJ...709.1195T} {709, 1195}

\bibitem[\protect\citeauthoryear{Treu et~al.,}{Treu
  et~al.}{2015}]{treu2015grism}
Treu T.,  et~al., 2015, \apj, 812, 114

\bibitem[\protect\citeauthoryear{{Trujillo}, {Conselice}, {Bundy}, {Cooper},
  {Eisenhardt}  \& {Ellis}}{{Trujillo} et~al.}{2007}]{Trujillo+07}
{Trujillo} I.,  {Conselice} C.~J.,  {Bundy} K.,  {Cooper} M.~C.,  {Eisenhardt}
  P.,   {Ellis} R.~S.,  2007, \mn@doi [\mnras]
  {10.1111/j.1365-2966.2007.12388.x}, \href
  {http://adsabs.harvard.edu/abs/2007MNRAS.382..109T} {382, 109}

\bibitem[\protect\citeauthoryear{Turner, Ostriker  \& Gott~III}{Turner
  et~al.}{1984}]{turner1984statistics}
Turner E.~L.,  Ostriker J.~P.,   Gott~III J.~R.,  1984, \apj, 284, 1

\bibitem[\protect\citeauthoryear{{Valentijn} et~al.,}{{Valentijn}
  et~al.}{2007}]{Valentijn2007}
{Valentijn} E.~A.,  et~al., 2007, in {Shaw} R.~A.,  {Hill} F.,   {Bell} D.~J.,
  eds,  Astronomical Society of the Pacific Conference Series Vol. 376,
  Astronomical Data Analysis Software and Systems XVI. p.~491 (\mn@eprint {}
  {astro-ph/0702189})

\bibitem[\protect\citeauthoryear{Van~der Walt, Sch{\"o}nberger, Nunez-Iglesias,
  Boulogne, Warner, Yager, Gouillart  \& Yu}{Van~der Walt
  et~al.}{2014}]{van2014scikit}
Van~der Walt S.,  Sch{\"o}nberger J.~L.,  Nunez-Iglesias J.,  Boulogne F.,
  Warner J.~D.,  Yager N.,  Gouillart E.,   Yu T.,  2014, PeerJ, 2, e453

\bibitem[\protect\citeauthoryear{{Vegetti} \& {Koopmans}}{{Vegetti} \&
  {Koopmans}}{2009}]{Vegetti2009}
{Vegetti} S.,  {Koopmans} L.~V.~E.,  2009, \mn@doi [\mnras]
  {10.1111/j.1365-2966.2009.15559.x}, \href
  {http://adsabs.harvard.edu/abs/2009MNRAS.400.1583V} {400, 1583}

\bibitem[\protect\citeauthoryear{{Verdugo} et~al.,}{{Verdugo}
  et~al.}{2014}]{Verdugo+14_SL2S}
{Verdugo} T.,  et~al., 2014, \mn@doi [\aap] {10.1051/0004-6361/201423696},
  \href {http://adsabs.harvard.edu/abs/2014A%26A...571A..65V} {571, A65}

\bibitem[\protect\citeauthoryear{{Whitmore}, {Lucas}, {McElroy},
  {Steiman-Cameron}, {Sackett}  \& {Olling}}{{Whitmore}
  et~al.}{1990}]{Whitmore1990}
{Whitmore} B.~C.,  {Lucas} R.~A.,  {McElroy} D.~B.,  {Steiman-Cameron} T.~Y.,
  {Sackett} P.~D.,   {Olling} R.~P.,  1990, \mn@doi [\aj] {10.1086/115614},
  \href {http://adsabs.harvard.edu/abs/1990AJ....100.1489W} {100, 1489}

\bibitem[\protect\citeauthoryear{{de Jong} et~al.,}{{de Jong}
  et~al.}{2015}]{deJong+15_KiDS_paperI}
{de Jong} J.~T.~A.,  et~al., 2015, \mn@doi [\aap]
  {10.1051/0004-6361/201526601}, \href
  {http://adsabs.harvard.edu/abs/2015A%26A...582A..62D} {582, A62}

\bibitem[\protect\citeauthoryear{{de Jong} et~al.,}{{de Jong}
  et~al.}{2017}]{deJong+17_KiDS_DR3}
{de Jong} J.~T.~A.,  et~al., 2017, preprint, \href
  {http://adsabs.harvard.edu/abs/2017arXiv170302991D} {} (\mn@eprint {arXiv}
  {1703.02991})

\bibitem[\protect\citeauthoryear{{de Vaucouleurs}}{{de
  Vaucouleurs}}{1948}]{deVauc48}
{de Vaucouleurs} G.,  1948, Annales d'Astrophysique, \href
  {http://adsabs.harvard.edu/abs/1948AnAp...11..247D} {11, 247}

\makeatother
\end{thebibliography}
\bsp

\appendix

\section{Neural Networks}\label{SECbackground} 
In this appendix we give a short introduction on the theory of Convolutional Neural Networks (CNN).

\subsection{Feed-forward neural network}
\begin{figure}
	\begin{center}
			\centering
			\includegraphics[width=0.3\textwidth]{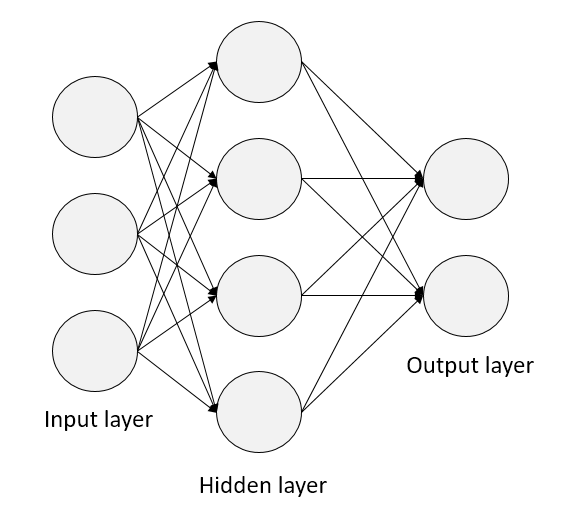}
			\caption{A schematic view of a feed forward neural network.}
			\label{FIGnn}
	\end{center}
\end{figure}

A feed-forward neural network, as the one shown schematically in \Fig\ref{FIGnn}, is a basic example of a deep-learning algorithm that can be schematized as an ensemble of connected units. The \textit{input layer} is the representation of a single element of the training set, where the units are the components of the data-point vector $\boldsymbol{x}=(x_1,x_2,...,x_n)$. Every unit in the \textit{hidden layer} performs the following transformation of its inputs

\begin{equation}
y=\sigma(\boldsymbol{w} \cdot \boldsymbol{x}+b)
\label{EQnnunit}
\end{equation}
where the vector $\boldsymbol{w} = (w_1,w_2,...,w_n)$ is called the \textit{weight vector} and the constant $b$ is the \textit{bias}. The non-linear activation function, $\sigma$ is often the Rectified linear unit (ReLU; \citealt{nair2010rectified})
\begin{equation}
\sigma(x)=\rm{max}(0,x),
\label{EQrelu}
\end{equation} 
or the sigmoid function  
\begin{equation}
\sigma(x)=\frac{1}{(1+e^{-x})}.
\label{EQrelu}
\end{equation} 
Hidden layers are stacked sequentially until the topmost, i.e. the \textit{output layer}, is reached. We want the output layer $\boldsymbol{y} = (y_1,y_2,...,y_n)^T$ to approximate the desired output $\boldsymbol{\hat{y}} =(\hat{y}_1,\hat{y}_2,...,\hat{y}_n)^T$. 
This is obtained by finding the weights and biases that minimize a chosen loss function $L(\boldsymbol{y},\boldsymbol{\hat{y}})$.
The minimization is most often done via the iterative process of \textit{gradient descent}. For each layer $l$ the weights and biases are updated in the following way
\begin{equation}
\begin{split}
&\boldsymbol{w}_l\rightarrow\boldsymbol{w'}_l=\boldsymbol{w}_l-\eta\frac{\partial L}{\partial \boldsymbol{w}_l}\\
&\boldsymbol{b}_l\rightarrow\boldsymbol{b'}_l=\boldsymbol{b}_l-\eta\frac{\partial L}{\partial \boldsymbol{b}_l}
\end{split}
\label{EQupdates}
\end{equation}
where $\eta$ is a constant called the \textit{learning rate}. The gradients are computed via the back-propagation algorithm \citep{backpropagation}.

\subsection{Convolutional Neural Network}

In Convolutional Neural Networks, the input data has a topological structure (e.g.\ an image) and is not presented as a vector but as a set of matrices $\boldsymbol{X}_k$ with $k=1,2,...,K$ 
(e.g., the R, G and B components of an image. In this case $K=3$). The main component of a CNN is the convolutional layer, which takes the inputs and, through a set of filters, produces a stack of \textit{feature maps} $\boldsymbol{Y_n}$ with $n$ equal to the number of filters. Every filter (also called kernel) in the convolutional layer produces a feature map through a convolution
\begin{equation}
\boldsymbol{Y}=\sigma\Bigg(\sum_{k=1}^K\boldsymbol{W}_k \ast \boldsymbol{X}_k+\boldsymbol{B}\Bigg),
\label{EQconvolution}
\end{equation}
where $\ast$ is the convolution operator, $\sigma$ is a non linear function as in eq. \eqref{EQnnunit}, $\boldsymbol{W}_k$ are the $K$ weight matrices with $k=1,2,...,K$, representing a filter with its bias given by the constant matrix $\boldsymbol{B}$. There are far fewer parameters to be determined in a convolutional layer as compared to a fully connected layer because, practically, we are replacing the dot product of \Eq\eqref{EQnnunit} with a convolution, and, in all the practical cases, the weight matrices have spatial dimensions much smaller than the input dimension (usually 3 by 3).

Convolutional layers are sequentially stacked such as the input of the deeper layers are the feature maps. In between then there are non-linear and other transformations (e.g. pooling). After the training is complete, in each layer we have a representation of the input data of increasing complexity (i.e. the different feature maps) each produced by a different filter which represents a particular feature learned during the training phase.   
The output layer of a CNN is the same as in a feed-forward neural network and can be preceded by one or more hidden fully-connected layers. Its function is to classify the last layer of feature maps created by the CNN, giving as output one ore more numbers which represent the outcome of the classification.    

\begin{figure*}
	\begin{center}
			\centering
			\includegraphics[height=6in,width=6in, keepaspectratio]{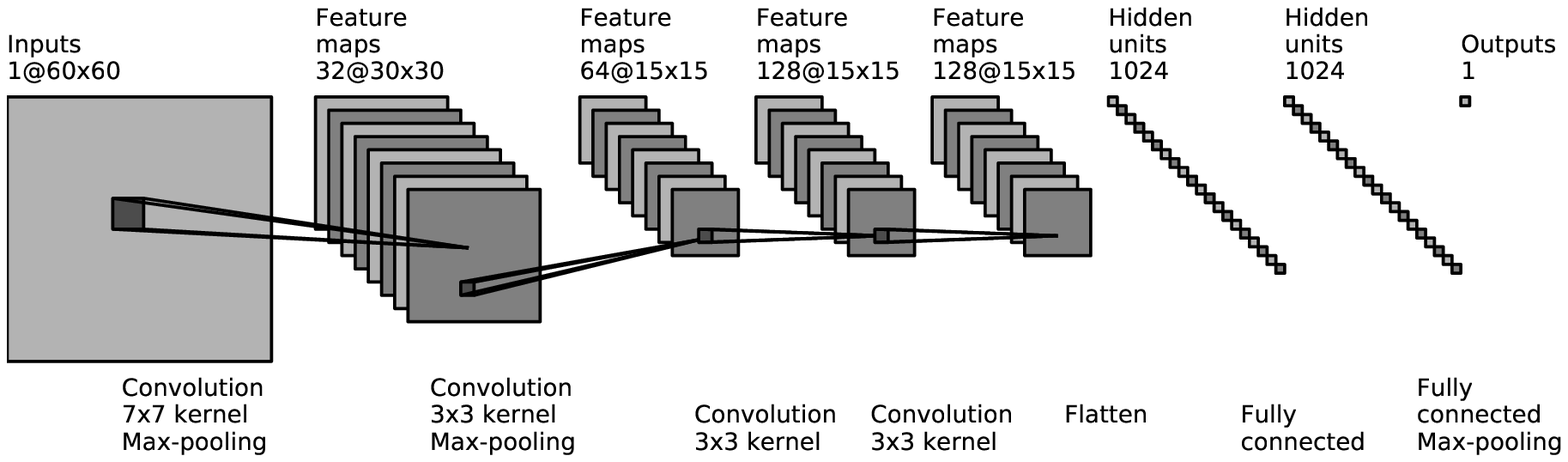} 
			\caption{Schematic view of the architecture of our CNN.}
			\label{figCNN}
	\end{center}
\end{figure*}

\begin{table*}
\caption{The table summarizes the characteristics of each layer of our CNN.}\label{TableCNN}
	\begin{center}
		\begin{tabular}{l*{6}{c}r}
Type              & Filters/Units & Filter size & Padding & Non-linearity & Initial weights  & Initial biases \\
\hline
Convolutional   & 32 & $7 \times 7$ & 3 & ReLU & HeNormal & 0  \\
Max-pooling     & - & $2 \times 2$ & - & - &  - & -  \\
Convolutional   & 64 & $3 \times 3$ & 1 & ReLU &  HeNormal & 0  \\
Max-pooling     & - & $2 \times 2$ & - & - &  - & -  \\
Convolutional   & 128 & $3 \times 3$ & 1 & ReLU &  HeNormal & 0  \\
Convolutional   & 128 & $3 \times 3$ & 1 & ReLU &  HeNormal & 0  \\
Fully connected & 1024 & - & - & ReLU &  HeNormal & 0  \\
Fully connected & 1024 & - & - & ReLU &  HeNormal & 0  \\
Max-pooling     & - & 1 & - & - &  - & -  \\
Fully connected & 1 & 2 & - & sigmoid &  HeNormal & 0  \\
		\end{tabular}
		\captionsetup{justification=centering}
		\label{tableparametersCNN}
	\end{center}
\end{table*}

\section{CNN implementation}\label{SEC:appendix}

Our CNN is implemented in Python 2.7 using the open-source libraries \textsc{Lasagne}\footnote{\href{http://github.com/Lasagne/Lasagne/}{\tt http://github.com/Lasagne/Lasagne/}} and \textsc{Theano}\footnote{\href{http://deeplearning.net/software/theano/}{\tt http://deeplearning.net/software/theano/}} \citep{theano}.  The training of the CNN is executed on a GeForce GTX 760 in parallel with the data augmentation performed on the CPU using the \textsc{scikit-image}\footnote{\href{http://scikit-image.org/}{\tt http://scikit-image.org/}} package \citep{van2014scikit}. The training time with this configuration takes about 2 hours. While the CNN takes about 20 minutes to classify the LRG sample.
In this Appendix we provide the technical details of the implementation and training of our CNN (\Secs\ref{arch} and \ref{train}).

\subsection{Network architecture}\label{arch}

In Fig. \ref{figCNN} and in \Tab\ref{TableCNN} we show the architecture of our CNN. ReLU (see Eq. \eqref{EQrelu}) is applied after each convolutional and fully connected layer.
The 60 by 60 input layer is followed by four convolutional layers with 32, 64, 128 and 128 filters, respectively. All the filters have 3 by 3 sizes except for the first convolutional layer which has a filter size of 7 by 7. In the convolutional kernels we use untied bias. To preserve the input volume through the convolution and not degrade the information at the borders of the input, we zero-pad the input of the convolutional layers with 3, 1, 1 and 1 pixels respectively. 
Max-pooling \citep{boureau2010learning} with a kernel size of 2 by 2 is used after the first and the second convolutional layer and after the second fully-connected layer with a 1D kernel size of 2. Max-pooling takes the maximum value in a connected set of elements of the feature maps. There are two main consequences of using Max-pooling: a) reducing the dimensionality of the data and the parameters to be estimated, b) teaching translational invariance to the network, because slightly shifted inputs will produce the same feature maps.
Two fully-connected layers of 1024 units follow the set of convolutional and max-pooling layers. Finally, we use a sigmoid non-linear output unit which gives a real number between 0 and 1 that represents the probability of being a strong gravitational lensing system. 
We use \textit{batch normalization} \citep{ioffe2015batch} before the non-linearity of each layer. A batch normalization layer operates on the inputs of the non-linearities normalizing the data in order to have zero mean and unit variance among the mini-batch. Then, the data is fed to a linear function with two learnable parameters that has the property to revert or modify the normalization.
The chosen architecture implies that our network has about 30 million of trainable parameters.

\subsection{Training}\label{train}
The network is trained by minimizing a loss function of the targets $t$ (1 for lenses and 0 for non-lenses) and the predictions $p$ (the output of the sigmoid unit of network). We use the binary cross-entropy, a common choice in two-class classification problems:
\begin{equation}
L = -t\log p - (1-t)\log(1-p)
\label{EQloss}
\end{equation}
The minimization is done via mini-batch stochastic gradient descent with \textit{ADAM} updates \citep{kingma2014adam}. The advantage of using \textit{ADAM} updates is the introduction of a friction term that mitigates the gradient momentum in order to reach a faster convergence. In addition, the updates have a per-parameter adaptation, i.e., they have a different effective learning rate for the different parameters depending on the gradient values.
We used a batch size of 600 images and perform 10000 gradient updates, which corresponds to six million examples. Each mini-batch is composed by 300 lens and 300 non-lens examples. After an initial exploration, we start with a learning rate of 0.004, decrease it to 0.0004 after three million training examples and to 0.00004 after 5.5 million training examples (the choice of the values for the learning rate has been fundamental for training successfully the network).
The weights of each filter are initialized, as discussed in \cite{he2015delving}, from a random normal distribution with variance ${2/n}$ where $n$ is the number of inputs of the unit. The initial values of the biases are set to zero. 
We use \textit{dropout} \citep{hinton2012improving} in the fully connected layers. Dropout consists of switching off units randomly during each update of the training phase. This has two main consequences: a) a speed-up of the training phase, because of the reduced number of parameters to be computed in the fully connected-layers of the CNN, b) reducing the possibility of over-fitting, since the network tends to learn features that better generalize the data.
We also use L2-norm regularization (see e.g., \citealt{ng2004feature}) with $\lambda={10}^{-4}$. The regularization adds to the loss function \Eq\eqref{EQloss} another factor given by the squared sum of all the weights times the factor $\lambda$. It has the property to let the network prefer to learn small diffuse weights penalizing the creation of peaky ones. In this way a classification, based on all the data coming from the input, tends to be promoted over one where the weights tend to consider only a subset of the input.

\begin{figure}
	\begin{center}
			\centering
			\includegraphics[height=3.5in,width=3.5in, keepaspectratio]{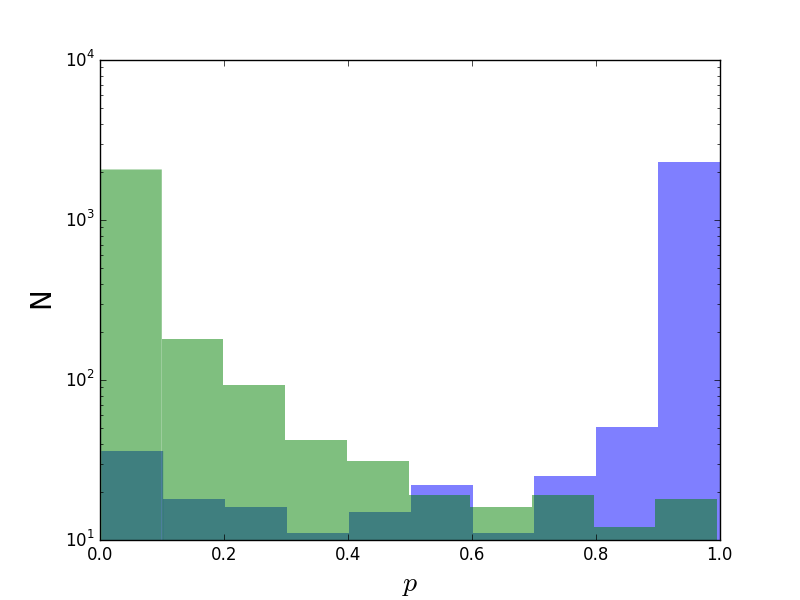} 
			\caption{The distribution of the network output for the validation set. The blue bars are the lens examples, while the green ones are the non-lens examples.}
			\label{hist_val}
	\end{center}
\end{figure}

\subsection{Analysis}\label{validation}

\begin{figure}
	\begin{center}
			\centering
			\includegraphics[height=3.3in,width=3.3in, keepaspectratio]{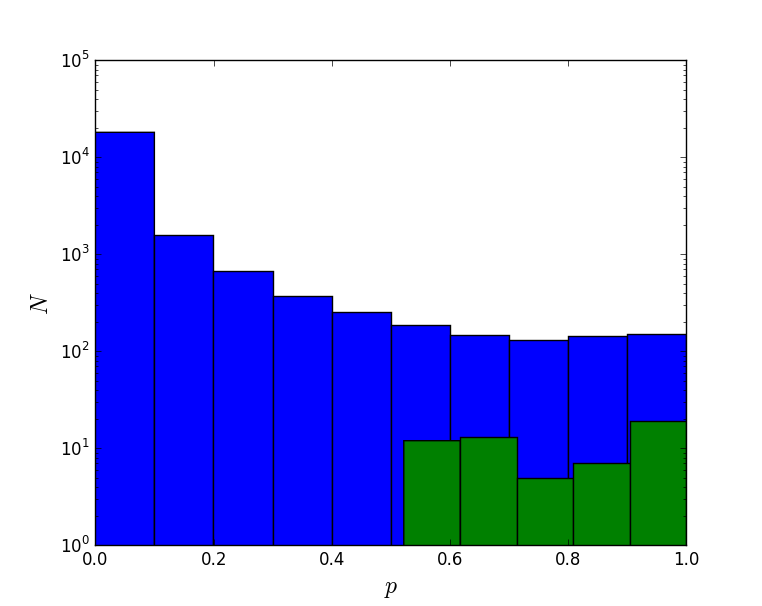} 
			\caption{The distribution of the network output for the full LRG sample (blue bars) and for the 56 candidates selected via visual inspection (green bars).}
			\label{FIGhist_real}
	\end{center}
\end{figure}

\begin{figure*}
   \centering
   \subfloat[(0.02 - 0.06)]{\includegraphics[width=33mm]{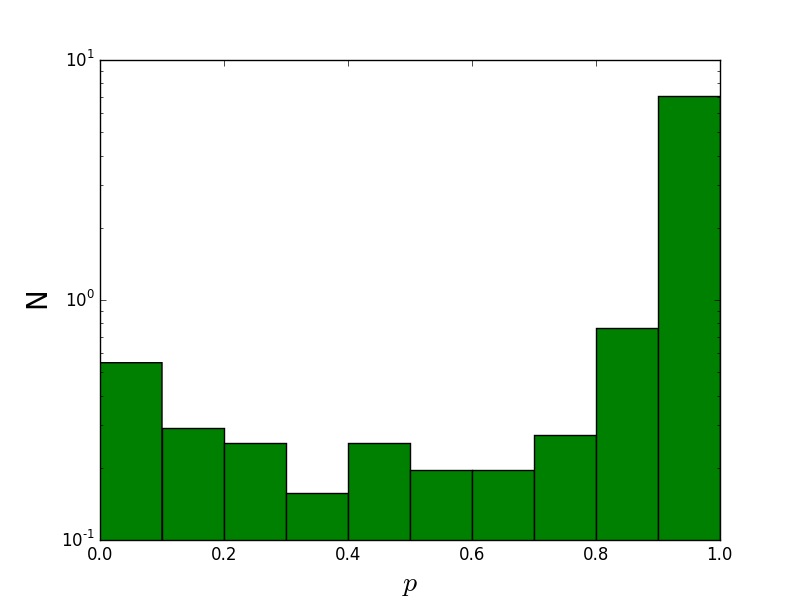}}\hspace{\fill}
   \subfloat[(0.06 - 0.09)]{\includegraphics[width=33mm]{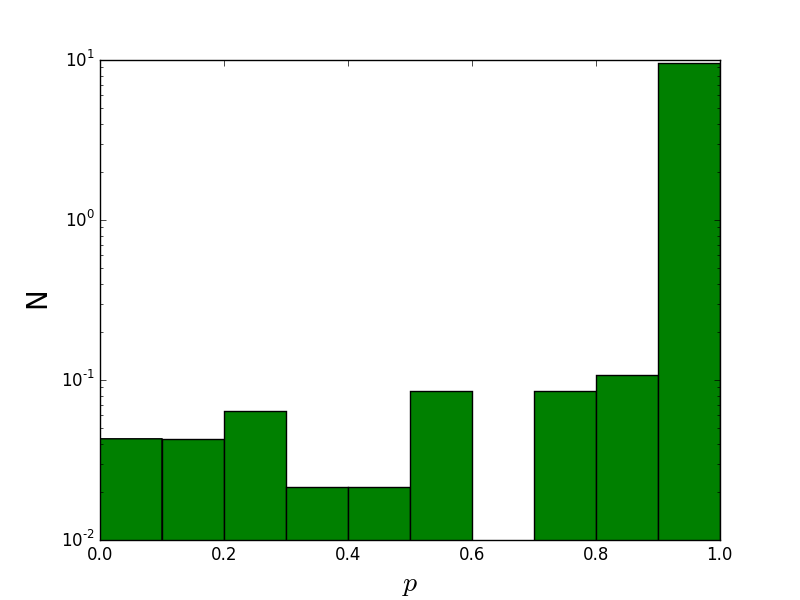}}\hspace{\fill}
   \subfloat[(0.09 - 0.13)]{\includegraphics[width=33mm]{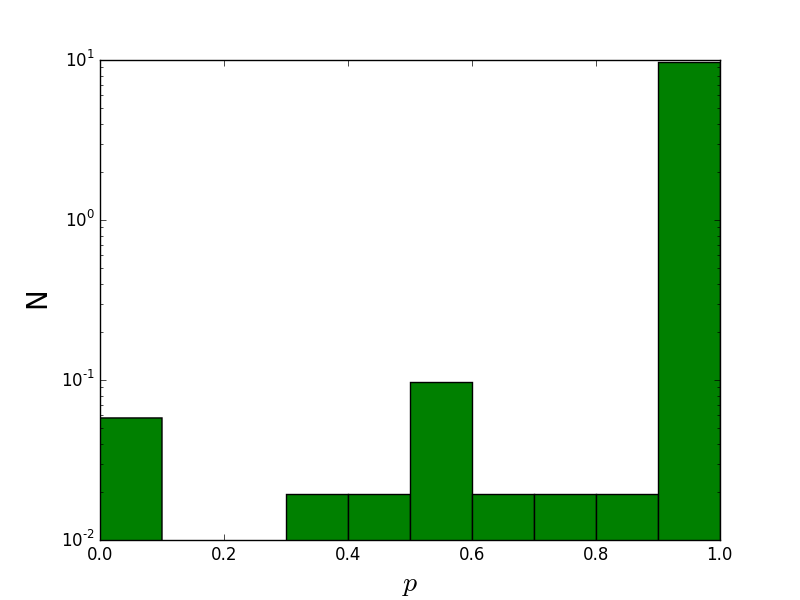}}\hspace{\fill}
   \subfloat[(0.13 - 0.16)]{\includegraphics[width=33mm]{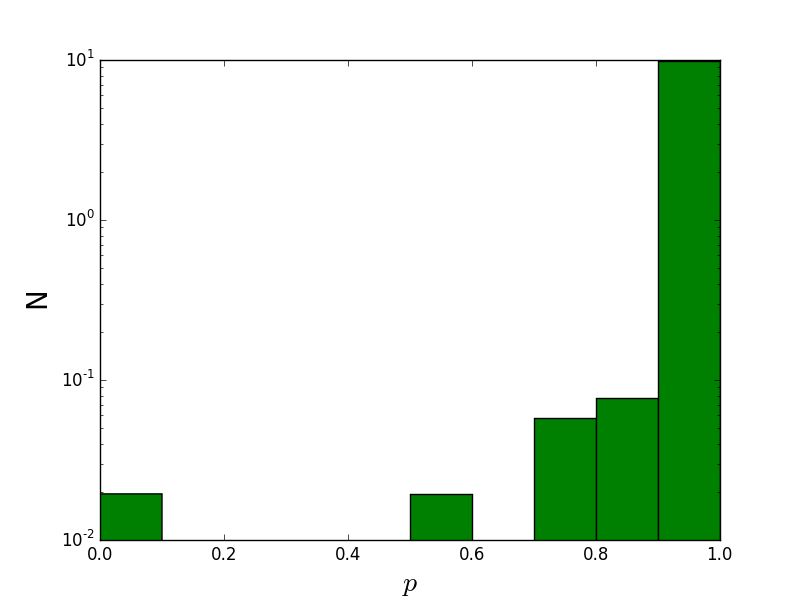}}\hspace{\fill}
   \subfloat[(0.16 - 0.20)]{\includegraphics[width=33mm]{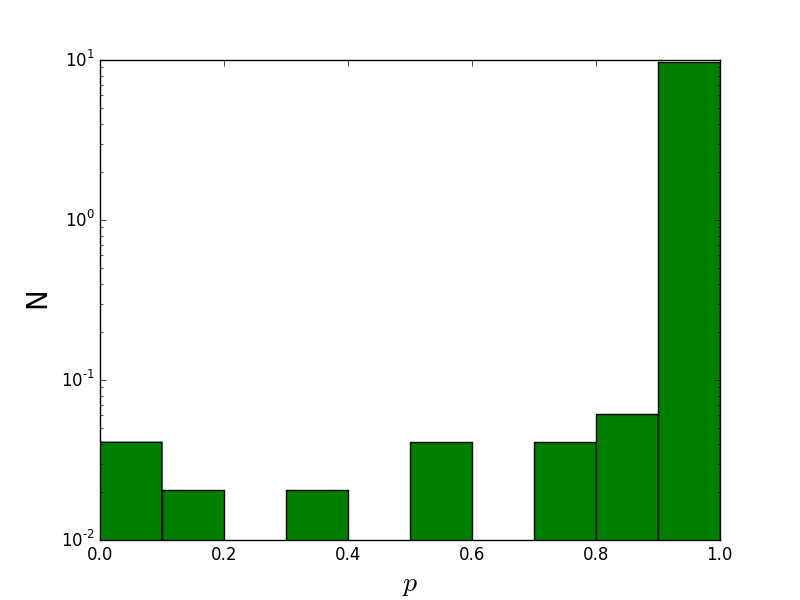}}\hspace{\fill}
   \subfloat[(1.4 - 2.12) arcsec]{\includegraphics[width=33mm]{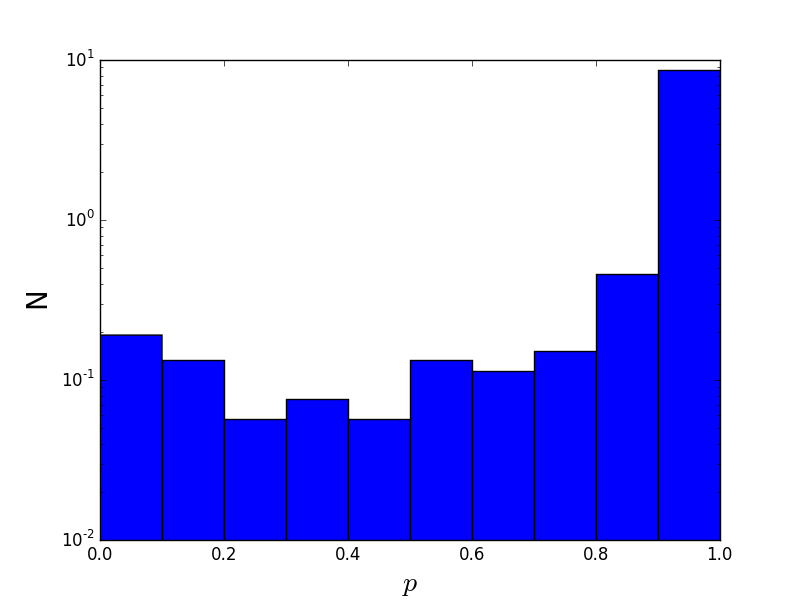}}\hspace{\fill}
   \subfloat[(2.12 - 2.84) arcsec]{\includegraphics[width=33mm]{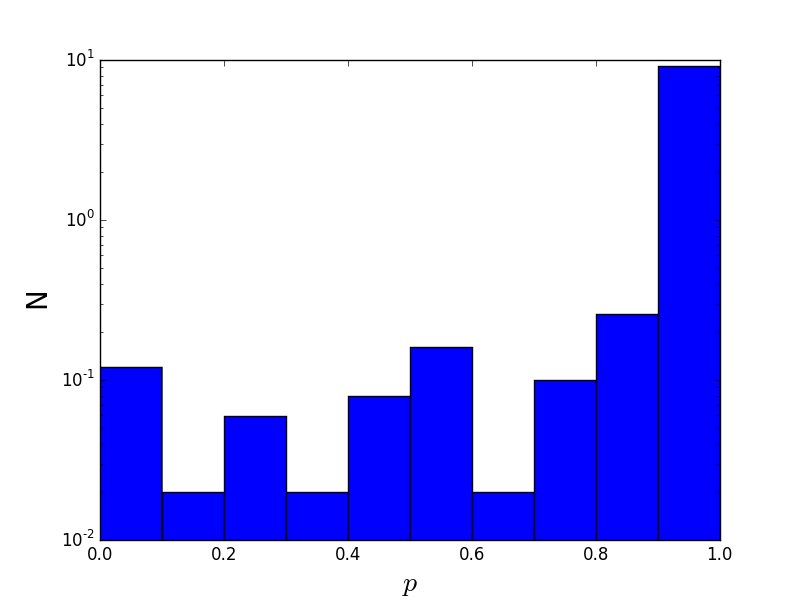}}\hspace{\fill}
   \subfloat[(2.84 - 3.56) arcsec]{\includegraphics[width=33mm]{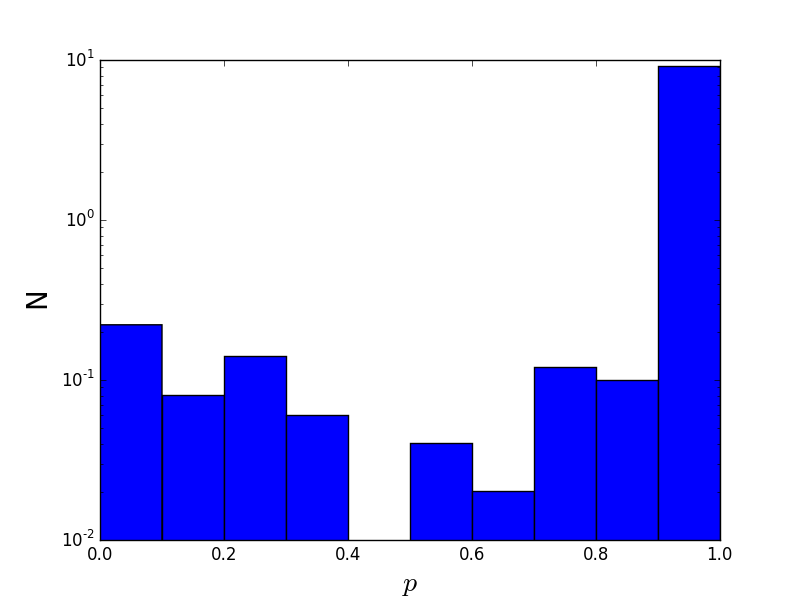}}\hspace{\fill}
   \subfloat[(3.56 - 4.28) arcsec]{\includegraphics[width=33mm]{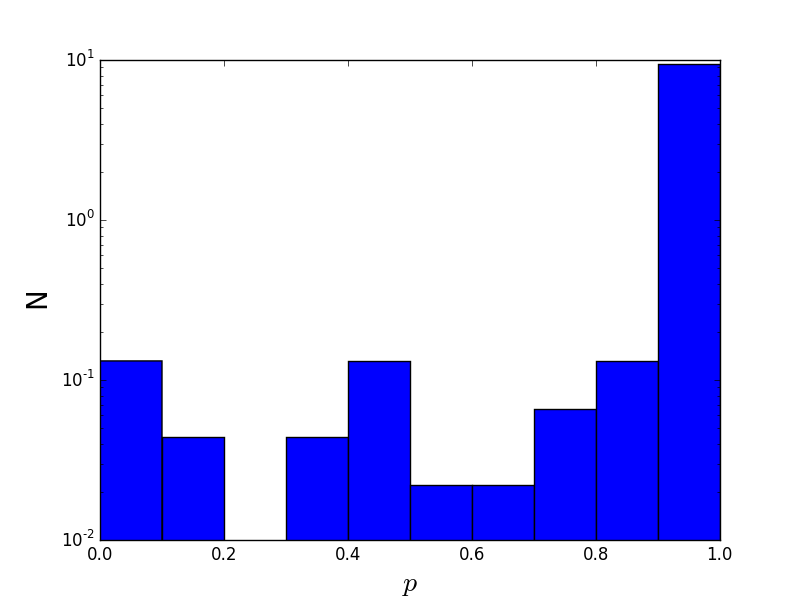}}\hspace{\fill}
   \subfloat[(4.28 - 5) arcsec]{\includegraphics[width=33mm]{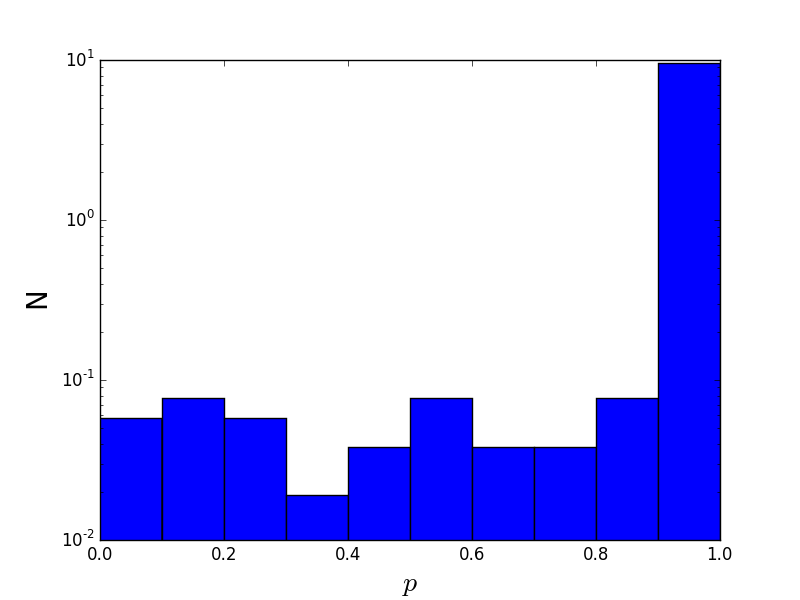}}\hspace{\fill}
   \subfloat[(1.1 - 11.1)]{\includegraphics[width=33mm]{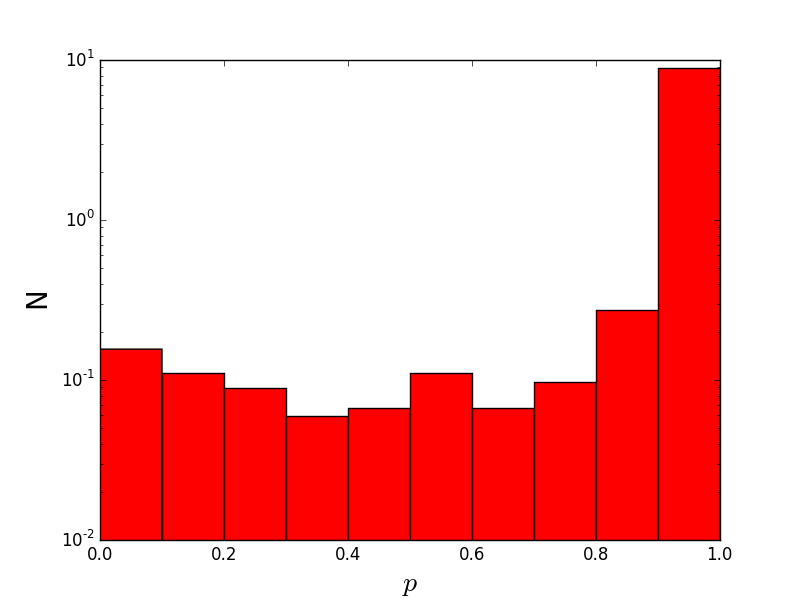}}\hspace{\fill}
   \subfloat[(11.1 - 21.1)]{\includegraphics[width=33mm]{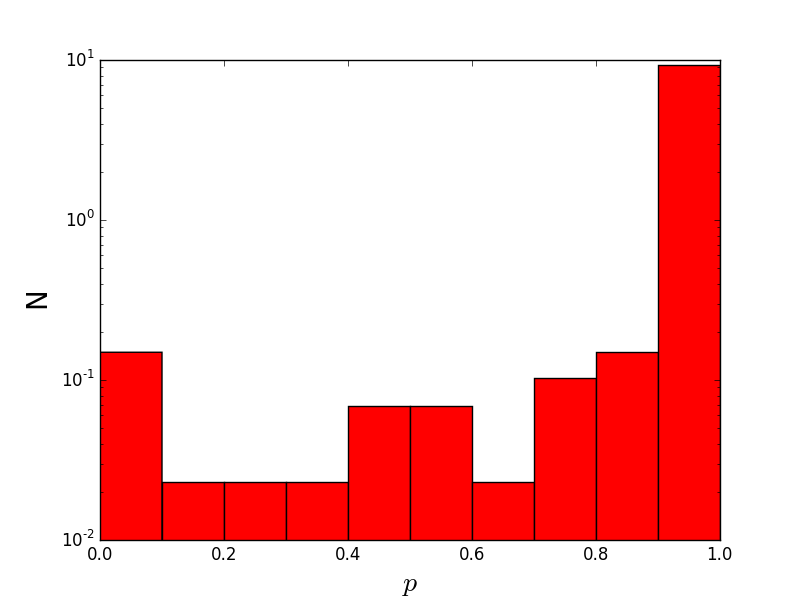}}\hspace{\fill}
   \subfloat[(21.1 - 31.1)]{\includegraphics[width=33mm]{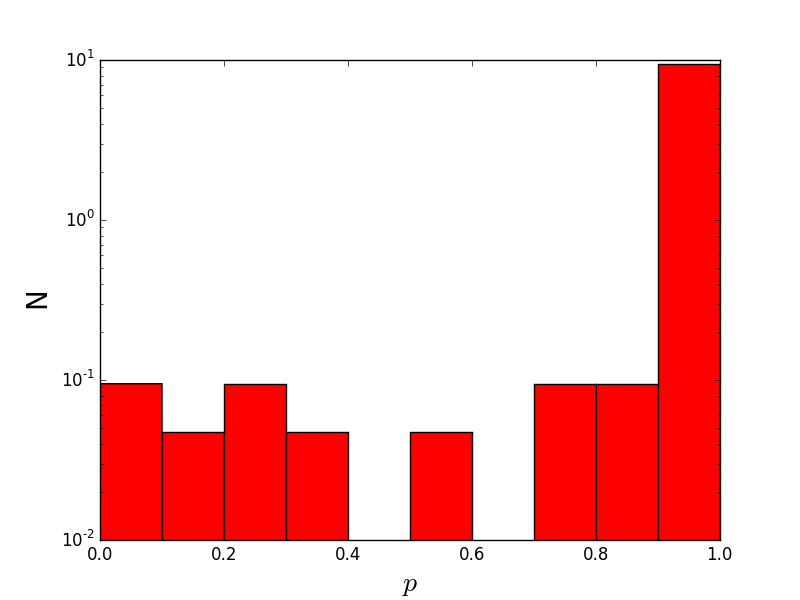}}\hspace{\fill}
   \subfloat[(31.1 - 71.1)]{\includegraphics[width=33mm]{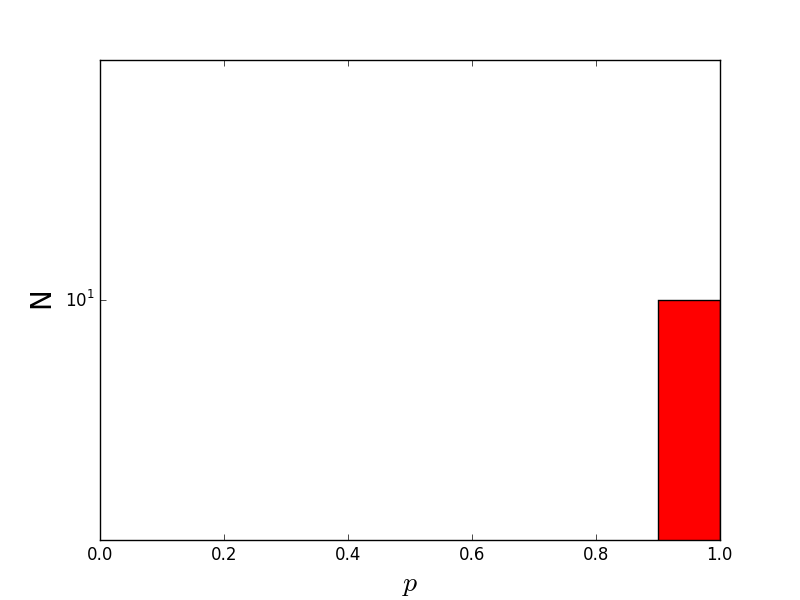}}\hspace{\fill}
      \subfloat[]{\includegraphics[width=33mm]{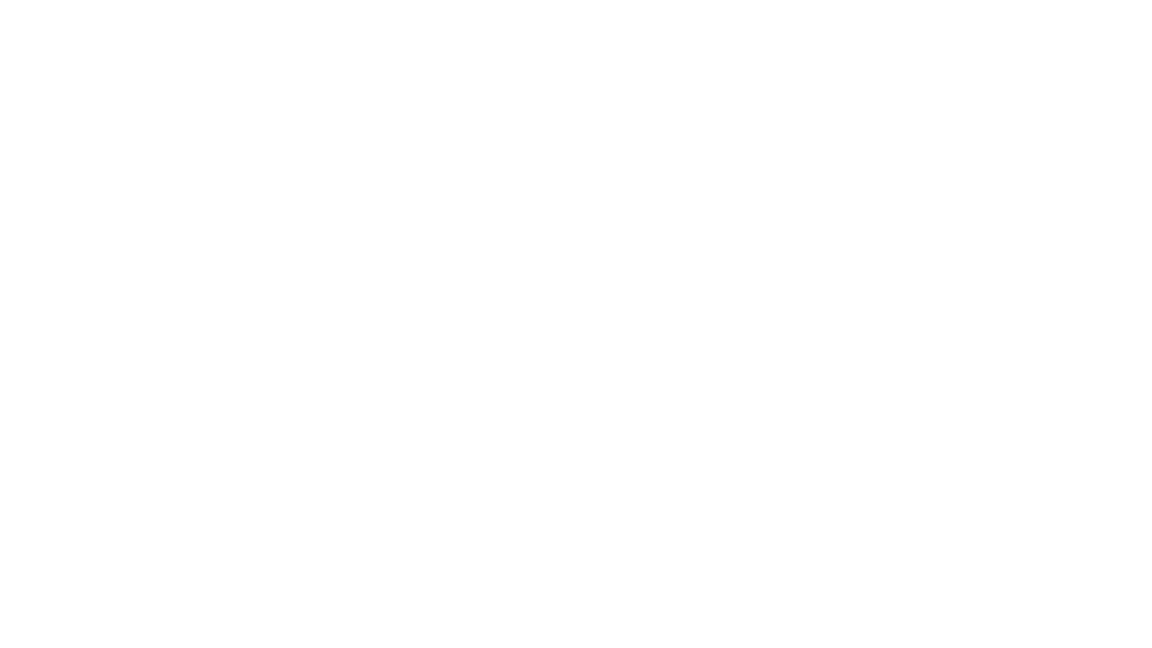}}\hspace{\fill}
   \caption{Distributions of the output of the CNN for different bins (shown in the parenthesis) of some parameters of the mock lensed sources (ratio between the maximum brightness of the lensed source and the lens in green, Einstein radius in blue, magnification in red).}
 \label{FIGhistograms}
 \end{figure*}

For monitoring the network during the training we build a fixed validation set composed of 5000 images (half lenses, half non-lenses) with the same prescriptions as summarized in \Sec\ref{sec:resume}.
At the end of the training the network reaches a 96\% accuracy for both the lens and non-lens examples. \Fig\ref{hist_val} shows the distribution of the network output $p$ for the validation set. For values of $p$ greater than 0.5 the lens examples start to be dominant in the distribution.
To check if there is any correlation between $p$ and the characteristics of the simulated mock sources, we investigate the $p$ distribution of the lens examples for different ranges of three parameters: Einstein radii, magnification and ratio between the peak brightness of the source and the lens. From \Fig\ref{FIGhistograms} one can see that the $p$ distributions  are skewed to higher values for the higher ranges of the parameters considered (larger, brighter and more magnified lens systems). This implies that the classification is more accurate for the training examples with higher magnification, higher Einstein radius and more luminous sources with respect to the KiDS galaxy. 
In \Fig\ref{FIGhist_real} we show the distribution of the network output of $p$-values for the full LRG sample and the 56 lens candidates (a sub-sample of the 761 candidates with $p>0.5$) . We do not retrieve a significant peaked distribution in the far end as for the validation set. This could be due to the intrinsic difference between real and simulated data. In addition, we have far fewer lenses in the LRG sample compared to the non-lenses. Moreover, the lenses in the validation set are uniformly distributed in the range of the parameters of \Tab\ref{TABLEmocksourceslens}. Thus, for a proper comparison, a validation set that reproduces the numbers and the distribution of the parameters for real lenses should be created.  

\captionsetup[subfigure]{labelformat=empty}

\section{r-band images of the candidates}\label{SEC:appendix2}

In \Fig\ref{FIGCandidatecutoutsrband} we show the \textit{r}-band images of the 56 candidates selected  through the visual inspection of \Sec\ref{SECvisual}. \textit{r}-band KiDS images have been used as the actual input of the CNN with the purpose of finding lens candidates.  

\captionsetup[subfigure]{labelformat=empty}
 \begin{figure*}
   \centering
   \subfloat[KSL427 (70)]{\includegraphics[width=38mm]{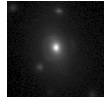}}\hspace{\fill}
   \subfloat[KSL317 (70)]{\includegraphics[width=38mm]{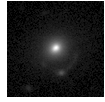}}\hspace{\fill}
   \subfloat[KSL103 (64)]{\includegraphics[width=38mm]{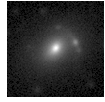}}\hspace{\fill}
   \subfloat[KSL627 (60)]{\includegraphics[width=38mm]{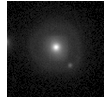}}\hspace{\fill}
   \subfloat[KSL040 (60)]{\includegraphics[width=38mm]{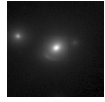}}\hspace{\fill}
   \subfloat[KSL327 (58)]{\includegraphics[width=38mm]{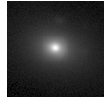}}\hspace{\fill}
   \subfloat[KSL376 (48)]{\includegraphics[width=38mm]{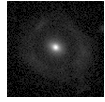}}\hspace{\fill}
   \subfloat[KSL086 (48)]{\includegraphics[width=38mm]{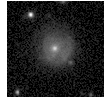}}\hspace{\fill}
   \subfloat[KSL469 (46)]{\includegraphics[width=38mm]{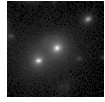}}\hspace{\fill}
   \subfloat[KSL351 (46)]{\includegraphics[width=38mm]{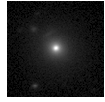}}\hspace{\fill}
   \subfloat[KSL713 (42)]{\includegraphics[width=38mm]{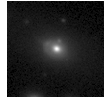}}\hspace{\fill}
   \subfloat[KSL328 (42)]{\includegraphics[width=38mm]{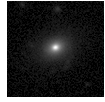}}\hspace{\fill}
   \subfloat[KSL228 (42)]{\includegraphics[width=38mm]{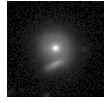}}\hspace{\fill}
   \subfloat[KSL411 (40)]{\includegraphics[width=38mm]{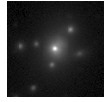}}\hspace{\fill}
   \subfloat[KSL070 (40)]{\includegraphics[width=38mm]{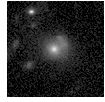}}\hspace{\fill}
   \subfloat[KSL543 (38)]{\includegraphics[width=38mm]{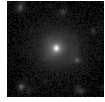}}
\caption{Square-root stretched KiDS \textit{r}-band images  of the 56 candidates selected through a visual inspection of the 761 CNN candidates (see \Sec\ref{SECvisual}). Each source is labelled by an internal ID followed by, in parenthesis, the visual classification score (70 points maximum). Each image is 20 by 20 arcsec.}
 \label{FIGCandidatecutoutsrband}
 \end{figure*}

\begin{figure*}
  \centering
  \subfloat[KSL664 (36)]{\includegraphics[width=38mm]{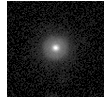}}\hspace{\fill}
  \subfloat[KSL106 (36)]{\includegraphics[width=38mm]{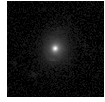}}\hspace{\fill}
  \subfloat[KSL415 (32)]{\includegraphics[width=38mm]{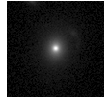}}\hspace{\fill}
  \subfloat[KSL388 (32)]{\includegraphics[width=38mm]{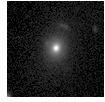}}\hspace{\fill}
  \subfloat[KSL337 (32)]{\includegraphics[width=38mm]{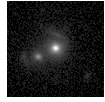}}\hspace{\fill}
  \subfloat[KSL220 (30)]{\includegraphics[width=38mm]{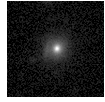}}\hspace{\fill}
  \subfloat[KSL603 (28)]{\includegraphics[width=38mm]{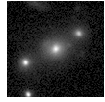}}\hspace{\fill}
  \subfloat[KSL601 (28)]{\includegraphics[width=38mm]{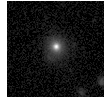}}\hspace{\fill}
  \subfloat[KSL737 (26)]{\includegraphics[width=38mm]{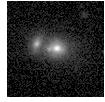}}\hspace{\fill}
  \subfloat[KSL669 (26)]{\includegraphics[width=38mm]{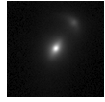}}\hspace{\fill}
  \subfloat[KSL450 (26)]{\includegraphics[width=38mm]{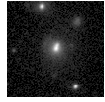}}\hspace{\fill}
  \subfloat[KSL436 (26)]{\includegraphics[width=38mm]{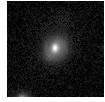}}\hspace{\fill}
  \subfloat[KSL233 (26)]{\includegraphics[width=38mm]{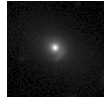}}\hspace{\fill}
  \subfloat[KSL231 (26)]{\includegraphics[width=38mm]{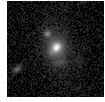}}\hspace{\fill}
  \subfloat[KSL101 (26)]{\includegraphics[width=38mm]{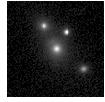}}\hspace{\fill}
  \subfloat[KSL094 (26)]{\includegraphics[width=38mm]{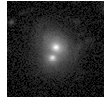}}\hspace{\fill}
  \subfloat[KSL707 (24)]{\includegraphics[width=38mm]{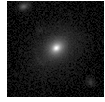}}\hspace{\fill}
  \subfloat[KSL565 (24)]{\includegraphics[width=38mm]{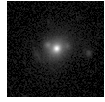}}\hspace{\fill}
  \subfloat[KSL335 (24)]{\includegraphics[width=38mm]{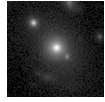}}\hspace{\fill}
  \subfloat[KSL197 (24)]{\includegraphics[width=38mm]{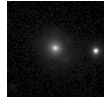}}
\contcaption{}                                        
\end{figure*}

\begin{figure*}
  \centering
  \subfloat[KSL134 (24)]{\includegraphics[width=38mm]{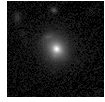}}\hspace{\fill}
  \subfloat[KSL620 (22)]{\includegraphics[width=38mm]{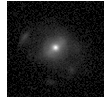}}\hspace{\fill}
  \subfloat[KSL606 (22)]{\includegraphics[width=38mm]{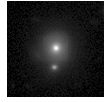}}\hspace{\fill}
  \subfloat[KSL434 (22)]{\includegraphics[width=38mm]{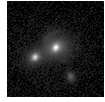}}\hspace{\fill}
  \subfloat[KSL421 (22)]{\includegraphics[width=38mm]{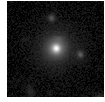}}\hspace{\fill}
  \subfloat[KSL046 (22)]{\includegraphics[width=38mm]{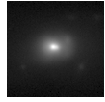}}\hspace{\fill}
  \subfloat[KSL013 (22)]{\includegraphics[width=38mm]{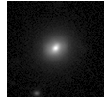}}\hspace{\fill}
  \subfloat[KSL686 (20)]{\includegraphics[width=38mm]{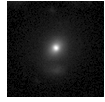}}\hspace{\fill}
  \subfloat[KSL674 (20)]{\includegraphics[width=38mm]{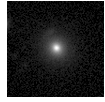}}\hspace{\fill}
  \subfloat[KSL670 (20)]{\includegraphics[width=38mm]{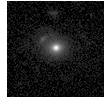}}\hspace{\fill}
  \subfloat[KSL564 (20)]{\includegraphics[width=38mm]{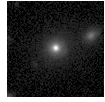}}\hspace{\fill}
  \subfloat[KSL516 (20)]{\includegraphics[width=38mm]{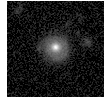}}\hspace{\fill}
  \subfloat[KSL465 (20)]{\includegraphics[width=38mm]{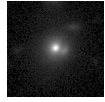}}\hspace{\fill}
  \subfloat[KSL463 (20)]{\includegraphics[width=38mm]{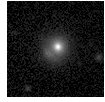}}\hspace{\fill}
  \subfloat[KSL342 (20)]{\includegraphics[width=38mm]{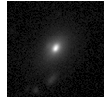}}\hspace{\fill}
  \subfloat[KSL322 (20)]{\includegraphics[width=38mm]{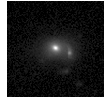}}\hspace{\fill}
  \subfloat[KSL278 (20)]{\includegraphics[width=38mm]{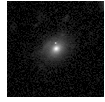}}\hspace{\fill}
  \subfloat[KSL178 (20)]{\includegraphics[width=38mm]{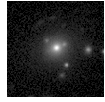}}\hspace{\fill}
  \subfloat[KSL159 (20)]{\includegraphics[width=38mm]{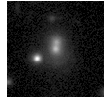}}\hspace{\fill}
  \subfloat[KSL535 (18)]{\includegraphics[width=38mm]{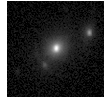}}
  \contcaption{}
\end{figure*}

\label{lastpage}
\end{document}